\documentclass[aps,prb,twocolumn,showpacs,superscriptaddress]{revtex4-1}
\usepackage{graphicx}
\usepackage{dcolumn}
\usepackage{bm}
\usepackage{hyperref}
\usepackage[utf8]{inputenc}
\usepackage[english]{babel}
\usepackage{units}
\usepackage{mathtools}
\hypersetup{colorlinks=true,citecolor={blue},linkcolor={blue},urlcolor={blue}}
\usepackage{amsmath}
\usepackage{empheq}
\usepackage{mathrsfs}
\usepackage{amssymb}
\usepackage{soul}
\setstcolor{red}

\begin{document}

\title{All-optical linear polarization engineering in single and coupled exciton-polariton condensates}

\author{I. Gnusov}
\email[I. Gnusov ]{Ivan.Gnusov@skoltech.ru}
\address{Skolkovo Institute of Science and Technology, Moscow, Territory of innovation center “Skolkovo”,
Bolshoy Boulevard 30, bld. 1, 121205, Russia.}

\author{H. Sigurdsson}

\email[H. Sigurdsson ]{h.sigurdsson@soton.ac.uk}
\address{Skolkovo Institute of Science and Technology, Moscow, Territory of innovation center “Skolkovo”,
Bolshoy Boulevard 30, bld. 1, 121205, Russia.}

\address{School of Physics and Astronomy, University of Southampton,  Southampton, SO17 1BJ, UK.}

\author{J. D. T{\"o}pfer}

\address{Skolkovo Institute of Science and Technology, Moscow, Territory of innovation center “Skolkovo”,
Bolshoy Boulevard 30, bld. 1, 121205, Russia.}

\author{S. Baryshev}
\address{Skolkovo Institute of Science and Technology, Moscow, Territory of innovation center “Skolkovo”,
Bolshoy Boulevard 30, bld. 1, 121205, Russia.}

\author{S. Alyatkin}
\address{Skolkovo Institute of Science and Technology, Moscow, Territory of innovation center “Skolkovo”,
Bolshoy Boulevard 30, bld. 1, 121205, Russia.}

\author{P. G. Lagoudakis}

\address{Skolkovo Institute of Science and Technology, Moscow, Territory of innovation center “Skolkovo”,
Bolshoy Boulevard 30, bld. 1, 121205, Russia.}
\address{School of Physics and Astronomy, University of Southampton,  Southampton, SO17 1BJ, UK.}

\begin{abstract}
We demonstrate all-optical linear polarization control of exciton-polariton condensates in anisotropic elliptical optical traps. The cavity inherent TE-TM splitting lifts the ground state spin degeneracy with emerging fine structure modes polarized linear parallel and perpendicular to the trap major axis with the condensate populating the latter. Our findings show a new type of polarization control with exciting perspectives in both spinoptronics and studies on extended systems of interacting nonlinear optical elements with anisotropic coupling strength and adjustable fine structure.
\end{abstract}
\maketitle

{\it Introduction.} --- Exciton-polaritons ({\it polaritons} hereafter) arise in the strong coupling regime between quantum well excitons and cavity photons in semiconductor microcavities~\cite{kavokin_microcavities_2007}. Being composite bosons, they can undergo a power-driven nonequilibrium phase transition into a highly coherent many-body state referred as polariton condensation~\cite{kasprzak_bose-einstein_2006}. An essential characteristic of polaritons is their spin projection ($\pm\hbar$) onto the growth axis of the cavity which corresponds to the right and
left circular polarizations of their photonic part.

The strong nonlinear nature of polaritons through their spin-anisotropic excitonic Coulomb interactions results in numerous intriguing spinor condensate properties. This includes spin bistability~\cite{pickup_optical_2018, Ohadi19, Sigurdsson_PRR2020} and multistability~\cite{Paraiso_Nature2010}, switches~\cite{Amo_NatPho2010, Cerna_NatComm2013}, optical spin Hall effect~\cite{leyder_observation_2007}, polarized solitons~\cite{Hivet_NatPhys2012, Sich_ACSPho2018} and vortices~\cite{Lagoudakis_Science2009, Donati_PNAS2016}, bifurcations~\cite{ohadi_spontaneous_2015}, and topological phases~\cite{Bleu_PRB2016, Sigurdsson_PRB2019}. Aforementioned opens great prospects for the utilization of the polariton spin degree of freedom in future spinoptronic technologies~\cite{pol_review, Liew_PhysE2011}. Different parts for future polariton based spin circuitry have already been realized~\cite{Amo_NatPho2010, Cerna_NatComm2013, Gao_APL2015, Dreismann_NatMat2016, askitopoulos_all-optical_2018} with some recent exciting theoretical proposals~\cite{sedov_LightScApp2019, Mandal_PRL2020}, but many challenges are still yet to be solved. Indeed, optical applications such as data communication or sensing benefit from precise control over a laser's polarization and modulation speeds, ideally using nonresonant excitation schemes like spin-VCSEL technologies~\cite{Ostermann_VCSELS2013, nature_vcsel_spinlasers, Drong_PRAppl2021}. In this spirit, a great deal of effort has been devoted to generating sources of linearly polarized light such as colloidal nanorods~\cite{Hu_Science2001}, materials with anisotropic optical properties~\cite{Wang_NatCommun2015}, quantum dots integrated into exotic structures~\cite{Lundskog_LScAppl2014}, and with optical parametric oscillators in the strong coupling regime~\cite{Krizhanovskii_PRB2006}.

Under nonresonant excitation in inorganic semiconductors, spin transfer from the pumping laser to the condensate is possible by creating a spin-imbalanced gain media for the circularly polarized polaritons (i.e., optical orientation of excitons) using an elliptically polarized beam~\cite{Ohadi19, Gnusov}. This allows generating polariton condensates of high degree of circular polarization aligned with the pump. However, in such systems the linearly polarized polariton modes experience isotropic gain, making it not possible to influence the linear polarization of the condensate under nonresonant excitation~\cite{ohadi_spontaneous_2012, baumberg_spontaneous_2008} except in the presence of cavity strain and birefringence~\cite{pinningfirst,Klopotowski_SSC2006, kasprzak_build_2007, balili_bose-einstein_2007, Read_PRB2009, Gnusov} or anisotropic confinement~\cite{ quantum_dot_weak_prb2019,  Klaas_PRB2019} inherent to the engineering of the cavity. The same also applies for VCSEL cavities, where the linear polarization of the emission is engineered by etching asymmetric masks~\cite{vcsel_mode_filter,pillar1998} or electrodes~\cite{vcsel_electrode}, heating~\cite{Pusch_APL2017}, or by applying mechanical stress~\cite{nature_vcsel_spinlasers}. 
Alternatively, in organic polaritonics, single-molecule Frenkel excitons can be excited by a linearly polarized pump co-aligned with their dipole moment with condensation into a mode with the
same linear polarization as the pump~\cite{Plumhof_organic}. However, control over both circular and linear polarization degrees of freedom in a polariton condensate through nonresonant all-optical means, instead of engineering specific cavity systems, remains elusive.

Here, we demonstrate in-situ optical engineering of the linear polarization in inorganic polariton condensates in a cavity with polarization-dependent reﬂectivity, or TE-TM splitting~\cite{Panzarini_PRB1999, leyder_observation_2007}. By spatially shaping the nonresonant excitation laser transverse profile into the form of an ellipse, we are able to fully control the direction of the condensate linear polarization.
Our elliptically shaped pumping profile induces an anisotropic in-plane trapping potential and gain media for the condensate. Such an excitation profile along with the cavity TE-TM splitting leads to condensation (lasing) into a mode of definite linear polarization parallel to the minor axis of the trap ellipse. The optical malleability of the trap geometry allows for non-invasive, yet deterministic, control over the linear polarization of the condensate by just utilizing the nonresonant excitation laser. Moreover, we investigate the effects of the anisotropic coupling mechanism between two spatially separated condensates and identify regions---as a function of coupling strength---of correlated high degree of random linear polarization between the condensates, and otherwise complete depolarization.
\begin{figure}[t!]
    \centering
    \includegraphics[width=1\columnwidth]{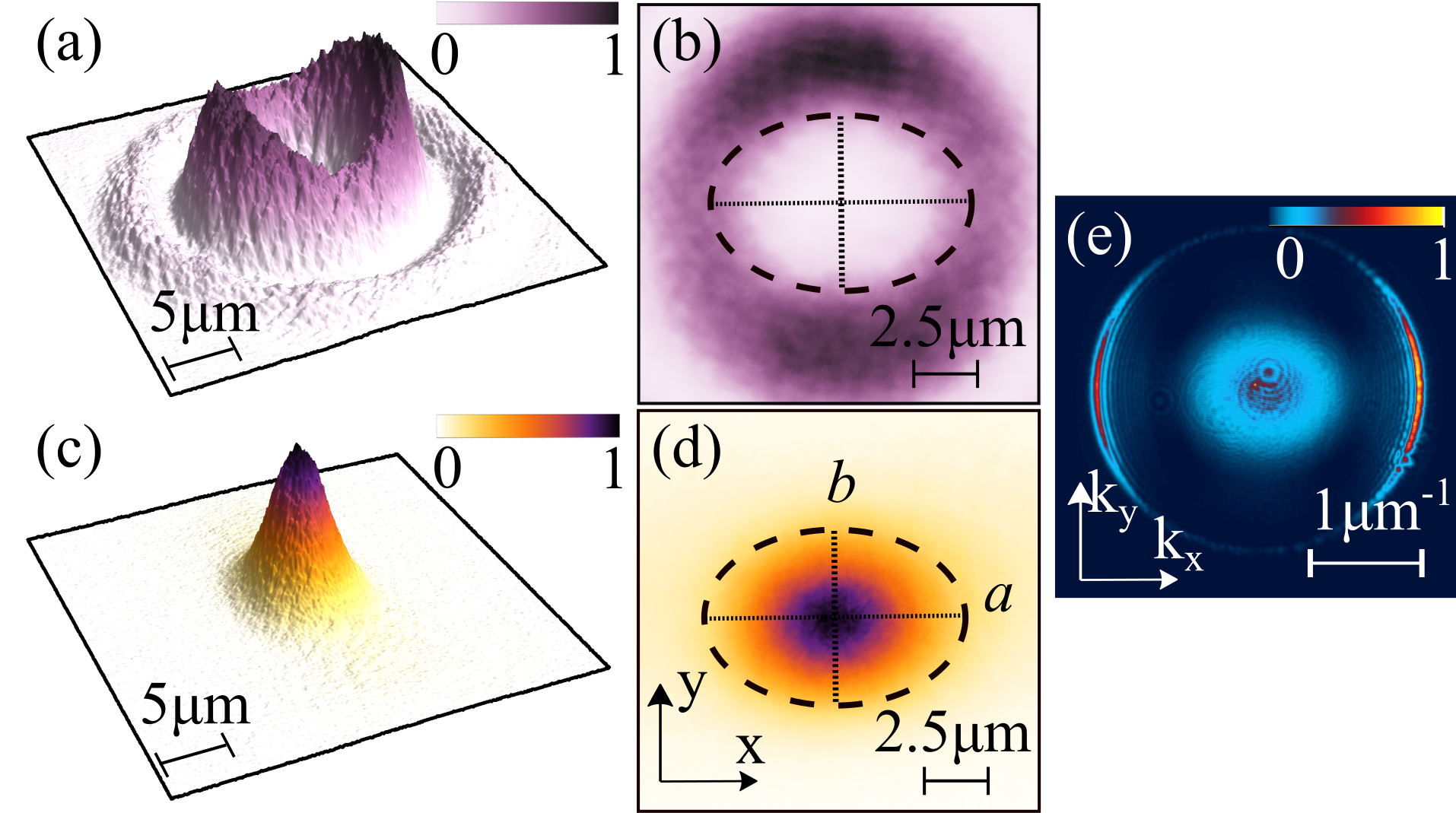}
    \caption{Spatial in-plane profiles of the (a,b) excitation laser and (c,d) condensate PL. The excitation laser induces a trapping potential with horizontal and vertical radii denoted $a,b$ respectively.  (e) Momentum distribution of the condensate PL. Panels (c,d,e) correspond to a condensate pumped twice above its condensation threshold (i.e., $P = 2P_{th}$).}
    \label{fig1}
\end{figure}

{\it Results.} --- Our experiments are conducted on an inorganic $2 \lambda$ GaAs/AlAs$_{0.98}$P$_{0.02}$ microcavity with embedded InGaAs quantum wells~\cite{cilibrizzi_polariton_2014}. The sample is excited nonresonantly  by a linearly polarized continuous wave (CW) laser ($\lambda=783.6$ nm). The optical excitation beam is chopped using an acousto-optic modulator to form 10 $\mu$s square pulses at 1 kHz repetition rate to diminish heating of the sample held at a temperature of 4 K. The exciton-cavity mode detuning is $-3$ meV. A reflective, liquid-crystal spatial light modulator (SLM) transforms the transverse profile of the pump laser beam to have an elliptically shaped confinement region [see Fig.~\ref{fig1}(a) and dashed ellipse in Fig.~\ref{fig1}(b)]. We investigate the sample PL in real [Fig.~\ref{fig1}(c,d)] and reciprocal [Fig.~\ref{fig1}(e)] space, and record the time- and space-averaged polarization of the PL by simultaneously detecting all polarization components~\cite{Gnusov}. Our results are independent on the angle of linear polarization of the pump laser [see Sec.~S1 in the Supplemental Information (SI)].

 The polariton condensate can be described by an order parameter written in the canonical spin-up and spin-down basis $\Psi = (\psi_+, \psi_-)^T$ corresponding to left- and right-circularly polarized condensate emission, respectively. It is then convenient to represent the condensate as a pseudospin on the Poincar\'{e} sphere corresponding to the Stokes vector (polarization) of the emitted light $\mathbf{S} = (S_1,S_2,S_3)^T = \langle \Psi^\dagger \boldsymbol{\hat{\sigma}}  \Psi \rangle / \langle \Psi^\dagger \Psi \rangle$ where $\boldsymbol{\hat{\sigma}}$ is the Pauli matrix vector. The PL is analyzed in terms of time-averaged Stokes components which are written as,
\begin{equation}
    S_1 = \frac{I_H - I_V}{I_H + I_V}, \
    S_2 = \frac{I_D - I_A}{I_D + I_A}, \
    S_3 = \frac{I_{\sigma^+} - I_{\sigma^-}}{I_{\sigma^+} + I_{\sigma^-}},
\end{equation}
where $I_{H,V,D,A,\sigma^+,\sigma^-}$ are the time-averaged intensities of horizontal, vertical, diagonal, antidiagonal, right- and left circular polarization projections of the emitted light.
\begin{figure}[t!]
    \centering
    \includegraphics[width=1\columnwidth]{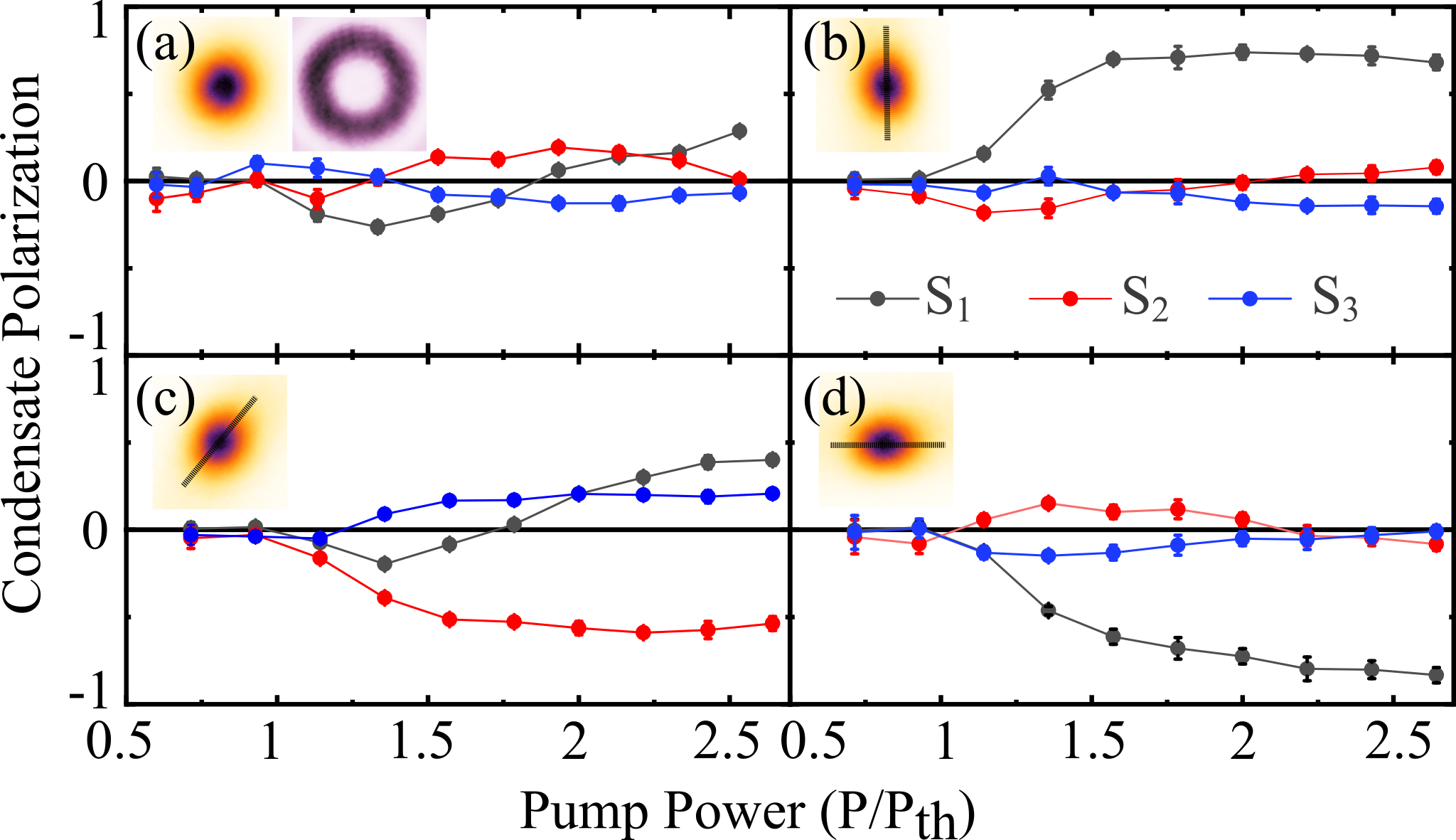}
    \caption{(a) Power dependence of the condensate $S_{1,2,3}$ Stokes parameters (black, red, blue markers) for the annular trap with resultant cyllindrically symmetric condensate profile. (b) Same but now for a trap/condensate with a major axis orientated at $90^{\circ}$, (c) $45^{\circ}$, and (d) $0^{\circ}$. Black lines in the insets depict the orientation of the trap major axis.}
    \label{fig2}
\end{figure}

We start by exciting with a symmetric ring-shaped pump profile  [see inset in Fig.~\ref{fig2}(a)], creating a two-dimensional trap for the polaritons and obtaining condensation with polaritons dominantly populating the trap ground state by ramping the pump power above the polariton condensation threshold denoted $P_{th}$. The optical trap is realized by the strong polariton repulsive interactions with the background laser-induced cloud of incoherent excitons which, in the mean field formalism, form a blueshifting potential onto the polaritons~\cite{askitopoulos_polariton_2013}, while at the same time providing gain to the condensate. Such an optical trapping technique has the advantage of reducing the overlap between the condensate and uncondensed excitons, minimizing detrimental dephasing effects. By scanning the excitation position with the ring-shaped pump profile we locate a spot on our sample with small degree of polarization $\text{DOP} = \sqrt{S_1^2 + S_2^2 + S_3^2}$ [see Fig.~\ref{fig2}(a)]. The small $S_{1,2}$ implies that the trap ground state is spin-degenerate such that from realization to realization random linear polarization builds up which averages out over many shots. The small $S_3$ component confirms that our laser excitation is (to a good degree) linearly polarized and doesn't break the spin parity symmetry of the system. We additionally investigate the condensate pumped with elliptical polarization in Sec.~S2 in the SI. 

We then transform the excitation profile to the one shown in Figs.~\ref{fig1}(a) and~\ref{fig1}(b). Non-uniform distribution of the intensity in the excitation leads to the formation of an elliptically shaped optical trap denoted by the dashed ellipse, squeezing the condensate as shown in Fig.~\ref{fig1}(d). 
We now observe a massive increase of the condensate's linear polarization components above 1.2$P_{th}$ at the same sample position. The direction of the linear polarization of the emission is found to follow the trap minor axis. Namely, for the vertically elongated condensate in Fig.~\ref{fig2}(b) we observe an increase of the $S_1$ Stokes component (horizontal polarization). The same effect is present for the horizontally and diagonally elongated condensates in Figs.~\ref{fig2}(c) and~\ref{fig2}(d). 

\begin{figure}[h]
    \centering
    \includegraphics[width=1\columnwidth]{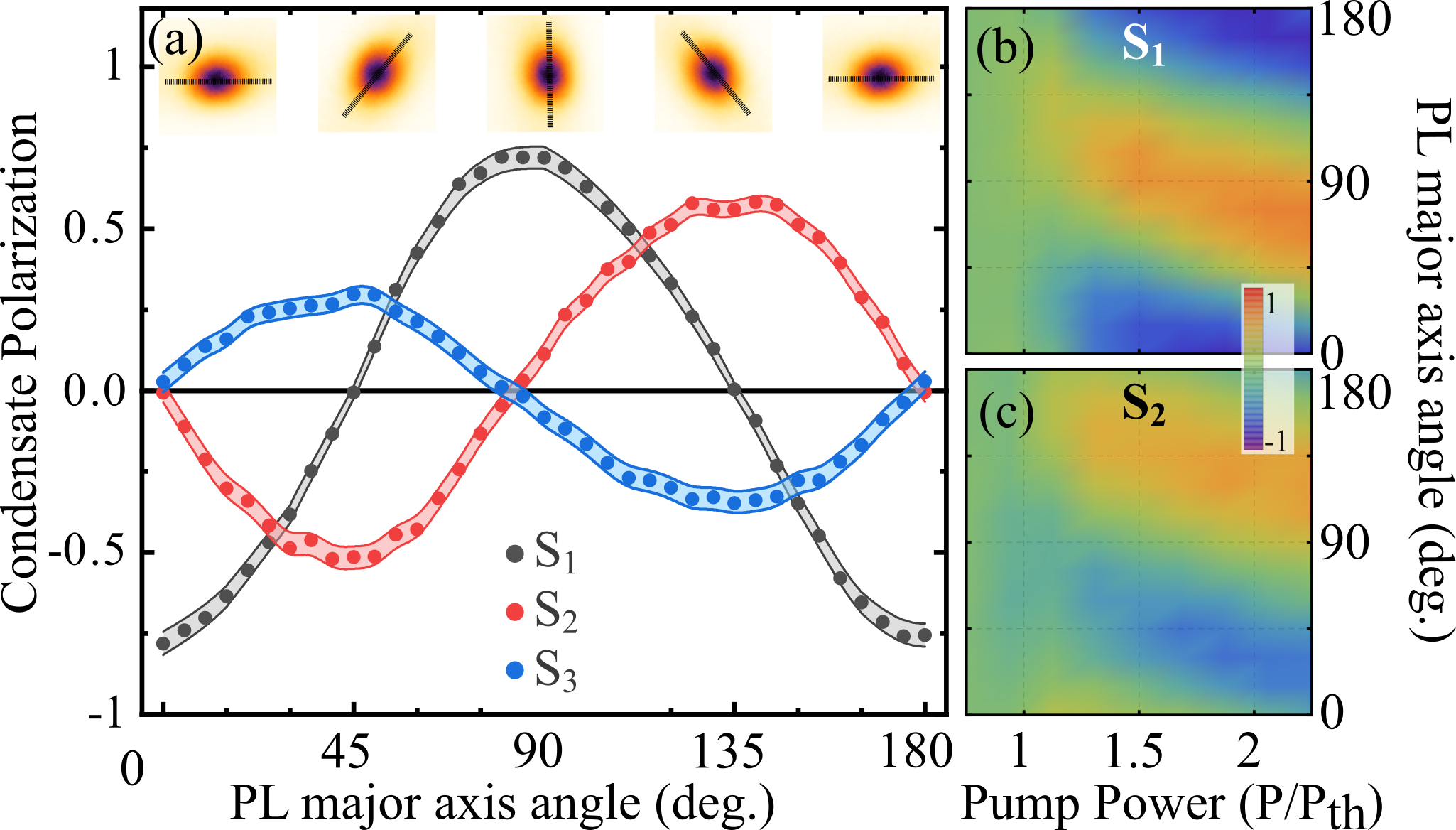}
    \caption{(a) Condensate Stokes parameters for different orientations of the condensate major axis in real space (insets) at $P=1.94P_{th}$.  Black lines depict the major axis of the trap. Colored regions show error of the measurement. (b) $S_1$ and (c) $S_2$ for varying pump powers and major axis orientation.}
    \label{fig3}
\end{figure}

By rotating the excitation profile with the SLM, we can engineer any desired linear polarization in the condensate. In Fig.~\ref{fig3}(a), we present the measured polarization components of the condensate as a function of the condensate major axis angle. We observe a continuous rotation of the condensate polarization close to the equatorial plane  of the Poincar\'{e} sphere following the minor axis of the trap. We also tested a different geometrical construction of the elliptical excitation profile with the same outcome (see Sec.~S4 in SI). We point out that $\text{DOP}<1$ appears from various depolarizing effects such as noise due to scattering from the incoherent reservoir to the condensate~\cite{Read_PRB2009}, polariton-polariton interactions in the condensate causing self-induced Larmor precessions~\cite{Ryzhov_PRR2020}, and mode competition~\cite{Redlich_NJP2016}. We also note that the finite $S_3$ component comes from optical elements in the detection path of our setup. 

Figures~\ref{fig3}(b) and~\ref{fig3}(c) show pump power and trap orientation dependence of the $S_{1,2}$ Stokes parameters. Interestingly, with increasing pump power we observe counterclockwise rotation of the pseudospin in the equatorial plane of the Poincar\'{e} sphere. The rotation is approximately $30^\circ$ between 1.2 and $2.2 P_{th}$. This effect appears due to a small amount of circular polarization in our pump which creates a spin-imbalanced trapping potential and gain media which acts as a complex population-dependent out-of-plane magnetic field $\boldsymbol{\Omega}_\perp$ that applies torque on the condensate pseudospin. This is confirmed through simulations using the generalised Gross-Pitaevskii equation (see Sec.~S10 in SI). Further analysis on this power dependent trend of the $S_{1,2}$ is beyond the scope of the current study.

\begin{figure}[t!]
    \centering
    \includegraphics[width=1\columnwidth]{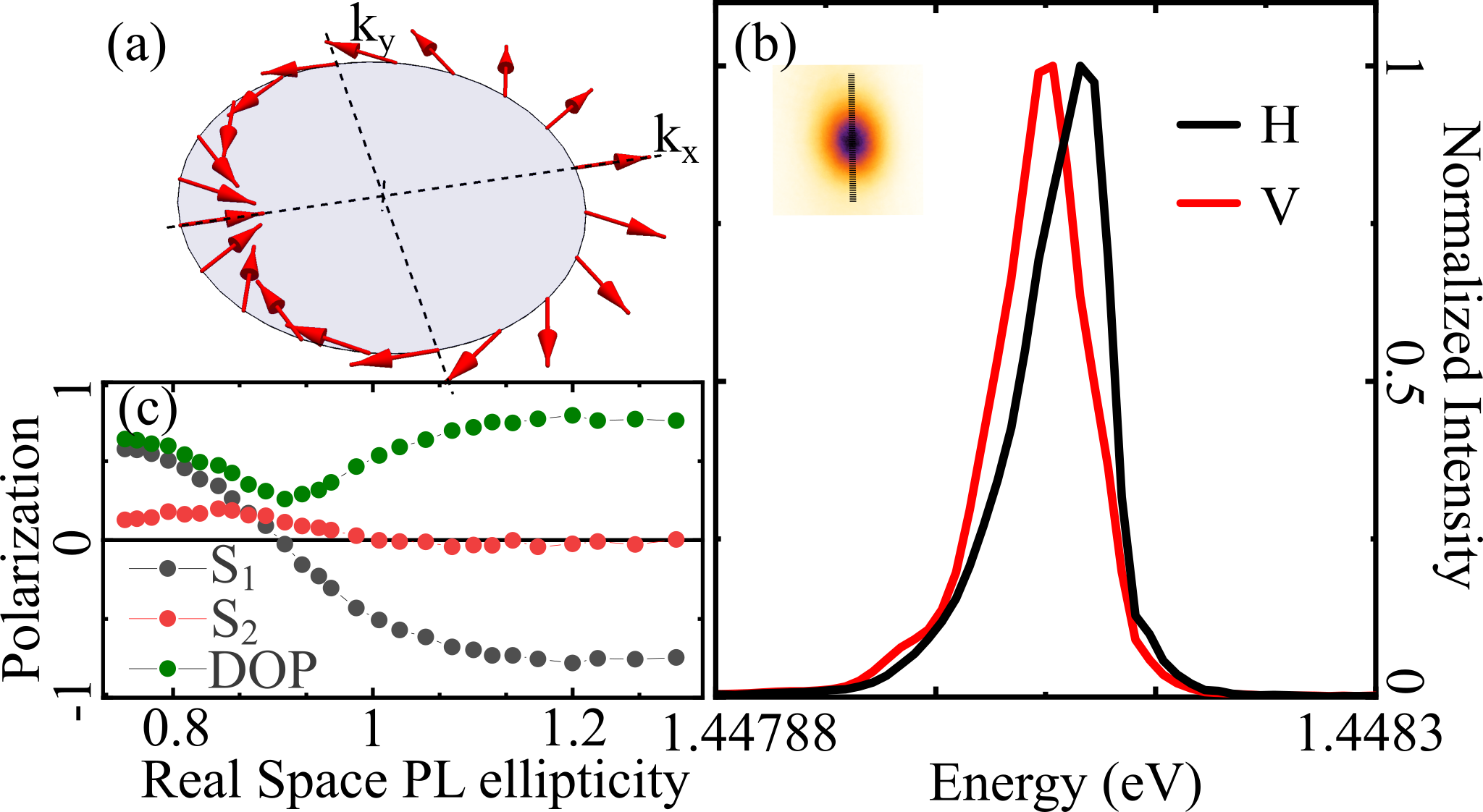}
    \caption {(a) Distribution of the in-plane effective magnetic field $\boldsymbol{\Omega}(\mathbf{k})$ (red arrows) in momentum space due to TE-TM splitting given by Eq.~\eqref{eq.Om}. (b) Horizontal (black) and vertical (red) polarization resolved normalized spectrum of the condensate emission at $k=0$ for a vertically elongated trap ($\theta_\text{min}=0$) favouring condensation into the horizontal mode. Splitting between levels is $\approx 20$ $\mu$eV. (c) Linear polarization components ($S_1$ and $S_2$) and DOP for different real-space ellipticities of the condensate. Data was taken at $P\approx 1.8P_{th}$}
    \label{fig4}
\end{figure}

Our observations can be interpreted in terms of photonic TE-TM splitting acting on the optically confined polaritons which, when the trap $V(\mathbf{r})$ has broken cylindrical symmetry, leads to fine structure splitting in the trap transverse modes. This determines a state of definite polarization which the polaritons condense into. In the noninteracting (linear) regime the polaritons obey the following Hamiltonian,
\begin{equation}
\hat{H} = \frac{\hbar^2 \boldsymbol{k}^2}{2m} - \boldsymbol{\hat{\sigma}} \cdot \boldsymbol{\Omega} + V(\mathbf{r}) - \frac{i \hbar \Gamma}{2},
 \end{equation}
where $m$ is the polariton mass, $\mathbf{k}=(k_x,k_y)$ is the in-plane cavity momentum, $\Gamma^{-1}$ is the polariton lifetime, and
\begin{equation} \label{eq.Om}
 \boldsymbol{\Omega} = \hbar^2 \Delta \left(
k_x^2 - k_y^2, \
2 k_x k_y, \
0
\right)^\text{T},
 \end{equation}
is the effective magnetic field [see Fig.~\ref{fig4}(a)] coming from the TE-TM splitting of strength $\Delta$~\cite{leyder_observation_2007}. In the considered case of an elliptical confinement, which we assume to be harmonic for simplicity $V(\mathbf{r}) = m(\omega_x^2 x^2 + \omega_y^2 y^2)/2$, the TE-TM splitting results in an effective magnetic field acting on the polariton pseudospin which splits the trap spin-levels. For the lowest (fundamental) harmonic state where most of the polaritons are collected this field can be written as follows (see Sec.~S5 in SI),
\begin{equation} \label{eq.Omtrap}
    \boldsymbol{\Omega}_\text{trap} = \frac{ \hbar m \Delta \delta\omega}{2} 
    \begin{pmatrix} 
    \cos{(2\theta_\text{min})} \\
    \sin{(2\theta_\text{min})} \\
    0 
    \end{pmatrix}.
\end{equation}
Here, $\theta_\text{min}$ is the angle of the trap minor axis from the horizontal, and $\delta \omega = |\omega_x - \omega_y| \propto |a^{-1} - b^{-1}|$ is the absolute difference between the trap oscillator frequencies along the major and the minor axis [Fig.~\ref{fig1}(d)]. We point out that $\Delta<0$ in our sample~\cite{Maragkou_OptLett2011} (see Sec.~S3 in SI). 

The direction of the effective magnetic field is controlled by the angle of our elliptical trap, $\theta_\text{min}$ which consequently rotates the condensate pseudospin in the equatorial plane of the Poincar\'{e} sphere such that it stabilizes antiparallel to the magnetic field $-\mathbf{S} \parallel \boldsymbol{\Omega}_\text{trap}$. This leads to smooth changes in the $S_{1,2}$ Stokes components of the emitted light as the trap rotates like shown in Fig.~\ref{fig3}.
The results of our experiment are accurately reproduced through a mean-field theory using a generalized Gross-Pitaevskii model describing the polariton condensate spinor order parameter $\Psi$ coupled with a background excitonic reservoir (see Sec.~S6 in SI). 

Interestingly, in a recent experiment~\cite{Gnusov} we observed condensation into the spin ground state of a circular trap, where the fine structure splitting originated from the cavity birefringence $\boldsymbol{\Omega}_\text{bir}(\mathbf{r})$. This meant that the condensate pseudospin stabilized parallel to the magnetic field $\mathbf{S} \parallel \boldsymbol{\Omega}_\text{bir}(\mathbf{r})$. In the current experiment however, we instead observe condensation into the excited spin state, i.e. antiparallel to the magnetic field $-\mathbf{S} \parallel \boldsymbol{\Omega}_\text{trap}$. This can be directly evidenced in Fig.~\ref{fig4}(b) where we show the normalized polarization-resolved spectrum of a vertically elongated trap which obtains a horizontally polarized condensate. The horizontal component is higher in energy in Fig.~\ref{fig4}(b), in agreement with Eq.~\eqref{eq.Omtrap}. 

Performing linear stability analysis on a Gross-Pitaevskii mean field model (see Sec.~S7 in SI) we determine that repulsive polariton-polariton interactions normally leads to condensation in the fine structure ground state~\cite{Read_PRB2009}. However, the additional presence of an uncondensed background of excitons (referred as the reservoir) contributes to an effective attractive mean-field interaction in the condensate~\cite{Estrecho_NatComm2018} which causes the ground state to become unstable, favouring condensation into the excited state as we observe in the current experiment. Another effect is the different penetration depths of the linearly polarized polariton modes (due to their different effective masses) into the excess gain region about the trap short axis. This leads to higher gain for the fine structure excited state which facilitates its condensation. Several parameters of the polariton system such as exciton-photon detuning, the quantum well material, and shape of the pump profile allow tuning from one stability regime to another which explains why some experiments show ground-state condensation~\cite{Gnusov} while other, like ours, show exited-state condensation~\cite{Maragkou_PRB2010}. We stress that regardless of whether system parameters favour condensation into the spin ground- or excited state of the optical trap, the main result of our study remains valid.
\begin{figure}[t]
    \centering
    \includegraphics[width=1\columnwidth]{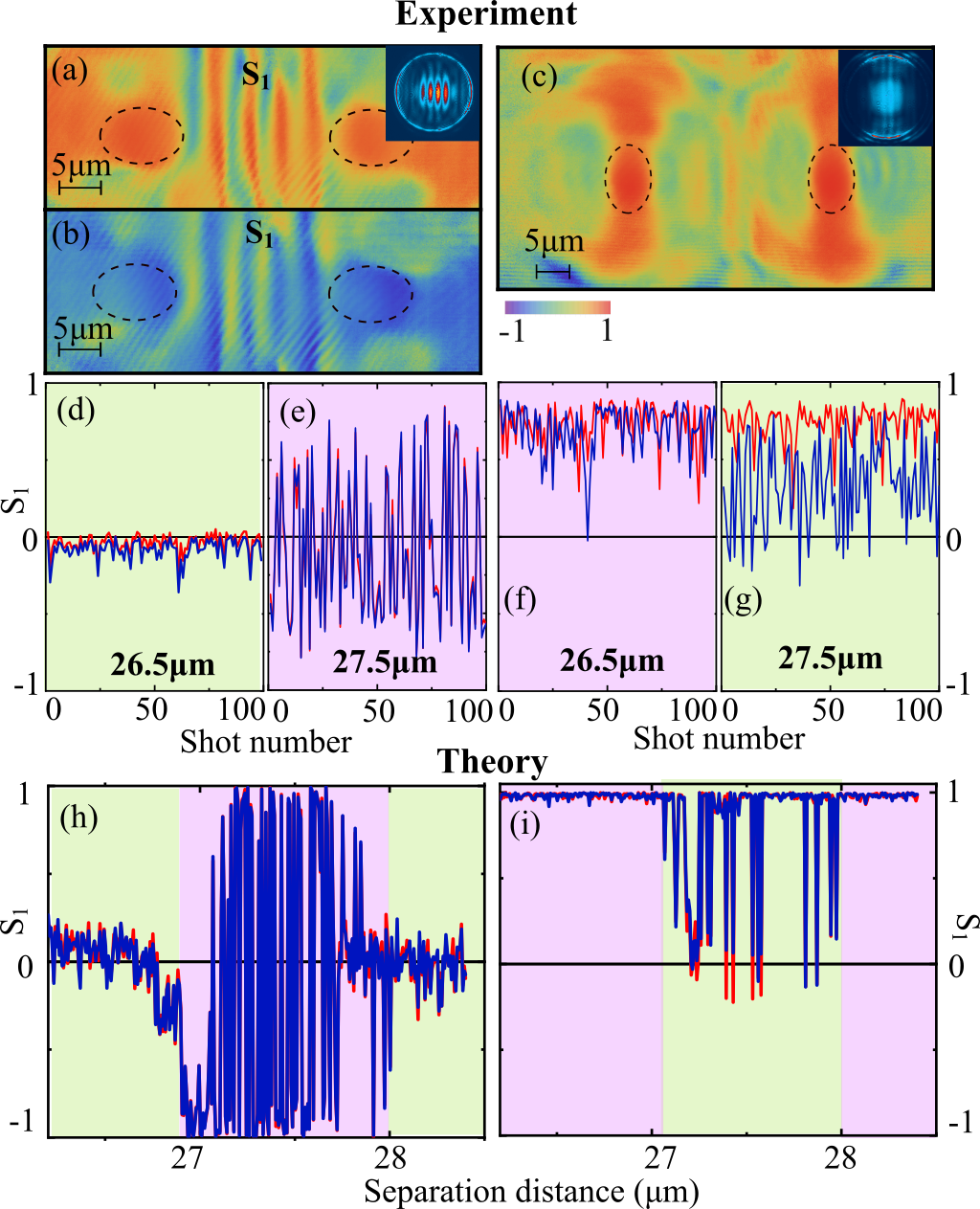}
    \caption {(a,b) Spatially resolved $S_1$ for two different realizations of coupled condensates oriented horizontally and separated by 27.5 $\mu$m, and (c) orientated vertically at 26.5 $\mu$m. Insets in (a) and (c) depict the corresponding $k$-space PL. (d,e) Show 100 time-integrated realizations (shots) of the $S_1$ for the left (blue) and right (red) condensate in the horizontal-horizontal major axis configuration and (f,g) in the vertical-vertical configuration. The experimental data is taken at $\approx$1.8 $P_{th}$. Distance dependence of the $S_1$ from Gross-Pitaevskii simulations corresponding to (h) horizontal-horizontal and (i) vertical-vertical major axes configuration. Each datapoint represents one shot (time averaged). Green and purple backgrounds in (d)-(i) illustrate regions of similar behaviour between experiment and theory.}
    \label{fig5}
\end{figure}

In Fig.~\ref{fig4}(c) we continuously change the trap ellipticity from a vertically elongated trap ($a<b$) to a horizontally elongated one ($a>b$). The PL ellipticity axis denotes the ratio of width of the PL along x- and y-axis [see Fig.~\ref{fig1}(d)]. The pseudospin of the condensate changes from horizontal to vertical polarization going through a low DOP regime. We stress that the data in Fig.~\ref{fig4}(c) is obtained at a different position of the cavity sample compared to Figs.~\ref{fig1}-\ref{fig3} which leads to finite $S_1$ at zero PL ellipticity ($a=b$) even though $\boldsymbol{\Omega}_\text{trap}=0$. This is because of local birefringence in the cavity mirrors giving rise to an additional static in-plane magnetic field $\boldsymbol{\Omega}_\text{bir}(\mathbf{r})$. Therefore, one needs to account for a net field  $\boldsymbol{\Omega}_\text{net} = \boldsymbol{\Omega}_\text{bir}(\mathbf{r}) + \boldsymbol{\Omega}_\text{trap}$ orientating the condensate pseudospin. The point of low DOP in Fig.~\ref{fig4}(c) corresponds then to near cancellation between the local birefringence and TE-TM splitting $\boldsymbol{\Omega}_\text{net}\approx0$.

{\it Coupled condensates. ---} Networks of coupled polariton condensates can be seen as an attractive platform to study the synchronization pheonomena in laser arrays, and to investigate the behaviour of complex nonequilibrium many-body systems and excitations in non-Hermitian lattices~\cite{ohadi_spin_2017, Mandal_PRL2020, Topfer_Optica2020, Pieczarka_arxiv2021}. Inspired by these studies, we create two identical, spatially separated, optical traps utilizing two SLMs resulting in the formation of two coupled condensates [Fig.~\ref{fig5}]. The trap anisotropy [see Fig\ref{fig1}(b)] allows polaritons to escape faster along its major axis[Fig\ref{fig1}(e)]. This leads to stronger coupling when the traps major axes are orientated longitudinally to the coupling direction, and weaker when orientated transverse (estimated as 3 times weaker from energy resolved spatial PL). This can be evidenced from the different visibility in the momentum space interference fringes (implying synchronization) [see insets in Figs.~\ref{fig5}(a) and~\ref{fig5}(c)].

Polarization resolving 100 quasi-CW 50 $\mu$s excitation shots, we observe distinct regimes depending on the condensates separation distance and orientation. For strongly coupled traps [Figs.~\ref{fig5}(a) and~\ref{fig5}(b)] at $26.5$ $\mu$m distance we observe zero DOP in each CW shot [Fig.~\ref{fig5}(d)] where the blue and red curves correspond to the left and right condensate. At a $27.5$ $\mu$m distance we now observe a strong $S_1$ component stochastically flipping from shot to shot [Fig.~\ref{fig5}(e)] with small $S_{2,3}$ (see S9 in SI)). Interestingly, the $S_1$ components of the condensates are almost perfectly correlated (Pearson correlation coefficient $\rho$  equals 0.99) which implies that they are strongly coupled. The linear polarization flipping suggests bistability in our system~\cite{Sigurdsson_PRR2020}, triggered by the spatial coupling mechanism. This interpretation is supported through Gross-Pitaevskii simulations on time-delay coupled spinor condensates presented in Fig.~\ref{fig5}(h) (see Sec.~S8 in SI). For weakly coupled traps [Fig.~\ref{fig5}(c)] we observe qualitatively different behaviour. Choosing again the same distances, we now see regimes of strong positive $S_1$ component [Fig.~\ref{fig5}(f)] and then semi-depolarized behaviour [Fig.~\ref{fig5}(g)]. Due to the weaker spatial coupling the condensates are no longer strongly correlated in their $S_1$ components ($\rho =0.5$ and $0.26$ respectively). We note that the different mean $S_1$ values in Fig.~\ref{fig5}(g) can be attributed to the position-dependent birefringence $\boldsymbol{\Omega}_\text{bir}(\mathbf{r})$. We reproduce the experiment from simulation [Fig.~\ref{fig5}(i)] by only decreasing the coupling strength by a factor of 3. We note that the ballistic (time-delayed) nature of the polariton condensate coupling~\cite{JulianTimedelay} distinguishes them from evanescently coupled quantum fluids. Indeed, the distance between the radiating condensates dictates their interference condition (in analogy to coupled laser systems) which---in our system---leads to distance-periodic appearance of the classified polarization regimes as seen in Fig.~\ref{fig5}(h) and~\ref{fig5}(i). Full dynamical trajectories from simulation, and a wider distance-power scan, are shown in Secs.~S8 and~S9 in the SI.

{\it Conclusion.} --- We have investigated the steady state polarization dynamics of a polariton condensate in an elliptically shaped trapping potential created through optical nonresonant linearly polarized injection. We have demonstrated that the polarization of the condensate is determined by the lifted spin-degeneracy of the trap levels due to the geometric ellipticity of the trap and inherent cavity TE-TM splitting. The condensate always forms in a higher energy spin state of the lowest trap level with a linear polarization that follows the minor axis of the trap ellipse. By rotating the excitation profile, we can rotate the condensate linear polarization around the equatorial plane of the Poincar\'{e} sphere. We have extended our system to coupled condensates, revealing rich physics of synchronization and desynchronization by tuning the condensate coupling strength through the optical trap anisotropy and/or spatial separation. Our results pave the way towards all-optical spin circuitry in spinoptronic applications, and coherent light sources with on-demand switchable linear polarization. 

The data presented in this paper are openly
available from the University of Southampton repository.

{\it Acknowledgements.} --- The authors acknowledge the support of the UK’s Engineering and Physical Sciences Research Council (grant EP/M025330/1 on Hybrid Polaritonics)   and by RFBR according to the research project No. 20-02-00919.

\setcounter{equation}{0}
\setcounter{figure}{0}
\setcounter{section}{0}
\renewcommand{\theequation}{S\arabic{equation}}
\renewcommand{\thefigure}{S\arabic{figure}}
\renewcommand{\thesection}{S\arabic{section}}
\onecolumngrid
\newpage
\vspace{1cm}
\begin{center}
\Large \textbf{Supplementary Information}
\end{center}

\section{Condensate polarization dependence on the linear polarization of the excitation laser}

\begin{figure}[b]
    \centering
    \includegraphics[width=0.6\columnwidth]{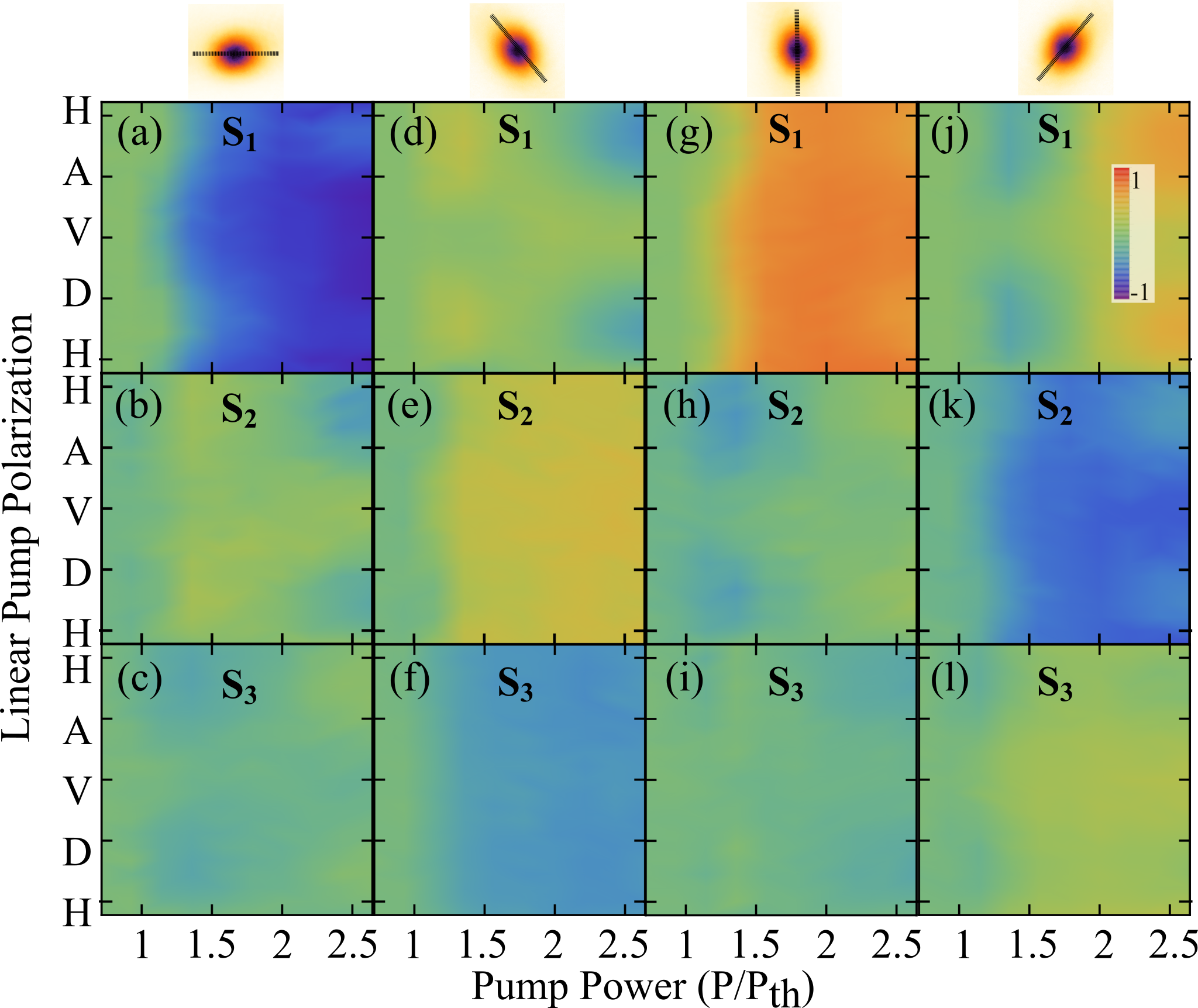}
    \caption{Condensate Stokes components for different pump powers and directions of linear polarization of the excitation laser. The labels H,A,V,D on the vertical axis denote horizontal, antidiagonal, vertical, and diagonal polarization respectively. The condensate PL is depicted on the top row with the black line denoting the trap major axis oriented at (a-c) 0, (d-f) -45, (g-i) 90, and (j-l) 45 degrees with respect to the cavity plane $x$-axis (horizontal direction).}
    \label{linpol}
\end{figure}

The experimental data presented in the main manuscript are acquired using a horizontally polarized pump laser which excites the optically trapped polariton condensate. In this supplemental section, we evidence that the linear polarization direction of the pump laser does not affect our presented results. In Fig.~\ref{linpol} we show the measured condensate photoluminescence (PL) Stokes components $S_{1,2,3}$ for varying power and linear polarization direction of the pump laser, the latter being controlled by a half-waveplate (HWP) in the excitation path. The four columns in Fig.~\ref{linpol} correspond to different spatial orientations of the elliptically shaped pump profile (i.e., the optical trap). Figures.~\ref{linpol}(a-c) are taken for $0^{\circ}$, (d-f) $-45^{\circ}$, (g-i) $90^{\circ}$, and (j-l) $45^{\circ}$ degrees of the trap ellipse major axis rotated counterclockwise from the horizontal direction (as defined in the main manuscript). We observe that the condensate polarization always dominantly follows the minor axis of the trap ellipse [see Fig.~\ref{linpol}(a),(e),(g), and~(k)].

The small amount of $S_3$ component emerging for diagonally oriented traps in Figs.~\ref{linpol}(f) and~\ref{linpol}(l) is due to optical elements in the detection path of our setup. For example, different reflectivities of the mirrors for $s$- and $p$-polarized light and small birefringence in the cryostat window glass. We have measured the effective retardance of the detection path in our setup to be $\approx $0.06$\pi$ at the condensate emission wavelength ($\approx 856$ nm).

\section{Condensate polarization dependence on the polarization ellipticity of the excitation laser}

\begin{figure}[h]
    \centering
    \includegraphics[width=0.6\columnwidth]{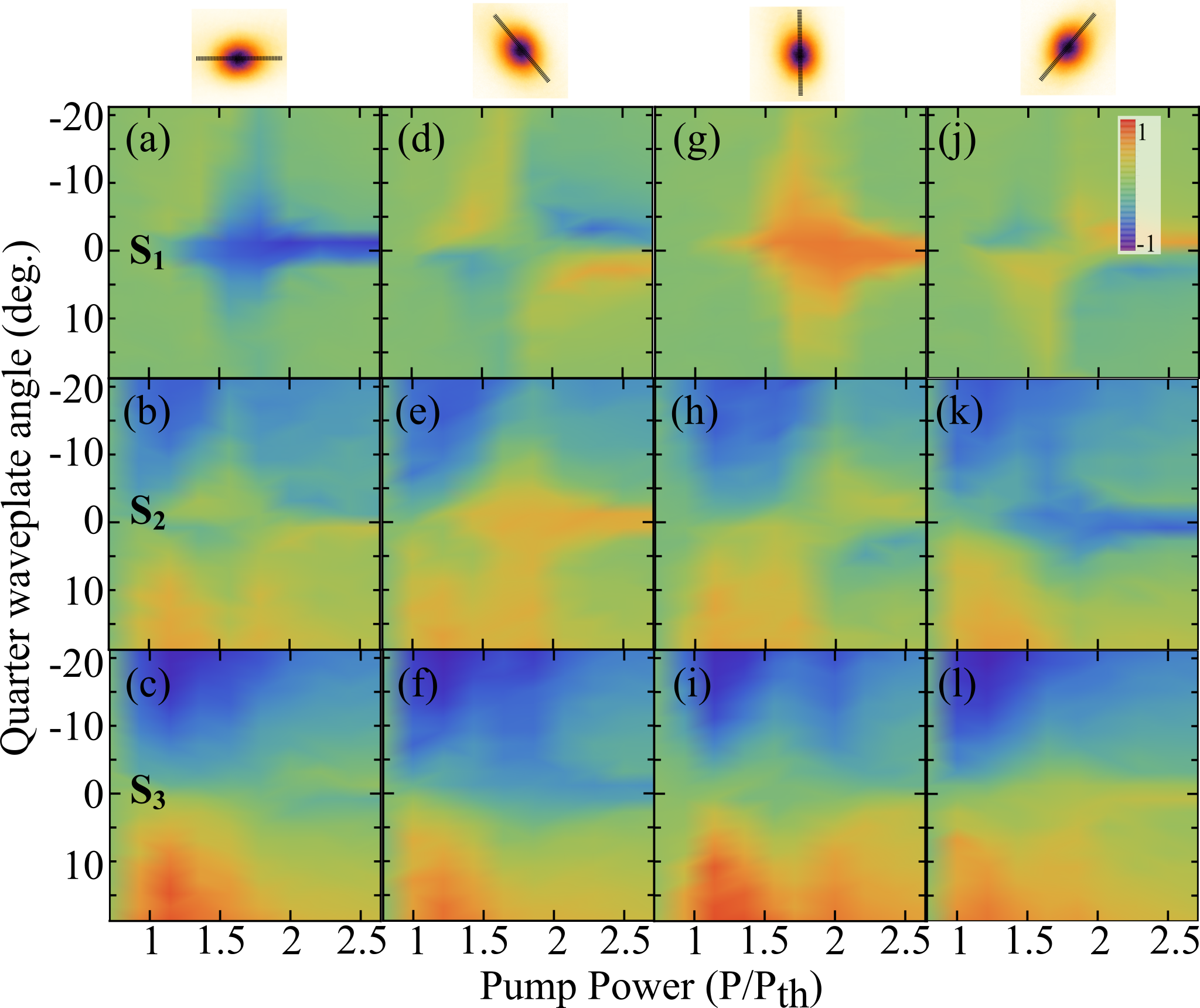}
    \caption{Condensate Stokes components for different pump powers and angles of the QWP (pump polarization ellipticity). The ellipticty of the pump can be written in normalized form as $S_3^\text{pump} = \sin{(2 \cdot \text{QWP})}$. The negative and positive values of the QWP angle correspond to left and right handedness of the circular polarization. The condensate PL is depicted on the top row with the black line denoting the trap major axis oriented at (a-c) 0, (d-f) -45, (g-i) 90, and (j-l) 45 degrees with respect to the cavity plane $x$-axis (horizontal direction).}
    \label{qwp}
\end{figure}

In this section we quantitatively investigate the dependence of the condensate polarization on the pump laser ellipticity. We install a quarter waveplate (QWP) in the excitation path so that by rotating the QWP, we can control the ellipticity and handedness of the excitation polarization. In Fig.~\ref{qwp} we show the measured condensate PL Stokes components $S_{1,2,3}$ depending on the pump polarization ellipticity and power. Overall, we obtain a similar behavior of the condensate polarization that was reported for annular optical traps~\cite{Gnusov_PRB2020s}.

As expected, circular polarization transfers to the condensate from our nonresonant excitation through the optical orientation of the background excitons feeding the condensate. It can also be seen that the linear polarization of the condensate is sensitive to the pump polarization ellipticity. This is effect is theoretically modeled and discussed further in Sec.~\ref{sec.S10}. In agreement with the findings presented in the main manuscript, when the pump is almost purely linearly polarized ($\text{QWP}\approx0$) we observe that the condensate aligns along the short axis of the optical trap [see e.g. blue coloured region in Fig.~\ref{qwp}(a)]. Our additional measurements in this supplemental section underline the richness of polarization regimes accessible in polariton condensates where, in this study, we have focused on anisotropic trapping conditions around $\text{QWP}\approx0$.

\section{TE-TM splitting}
\begin{figure}
    \centering
    \includegraphics[width=0.7\columnwidth]{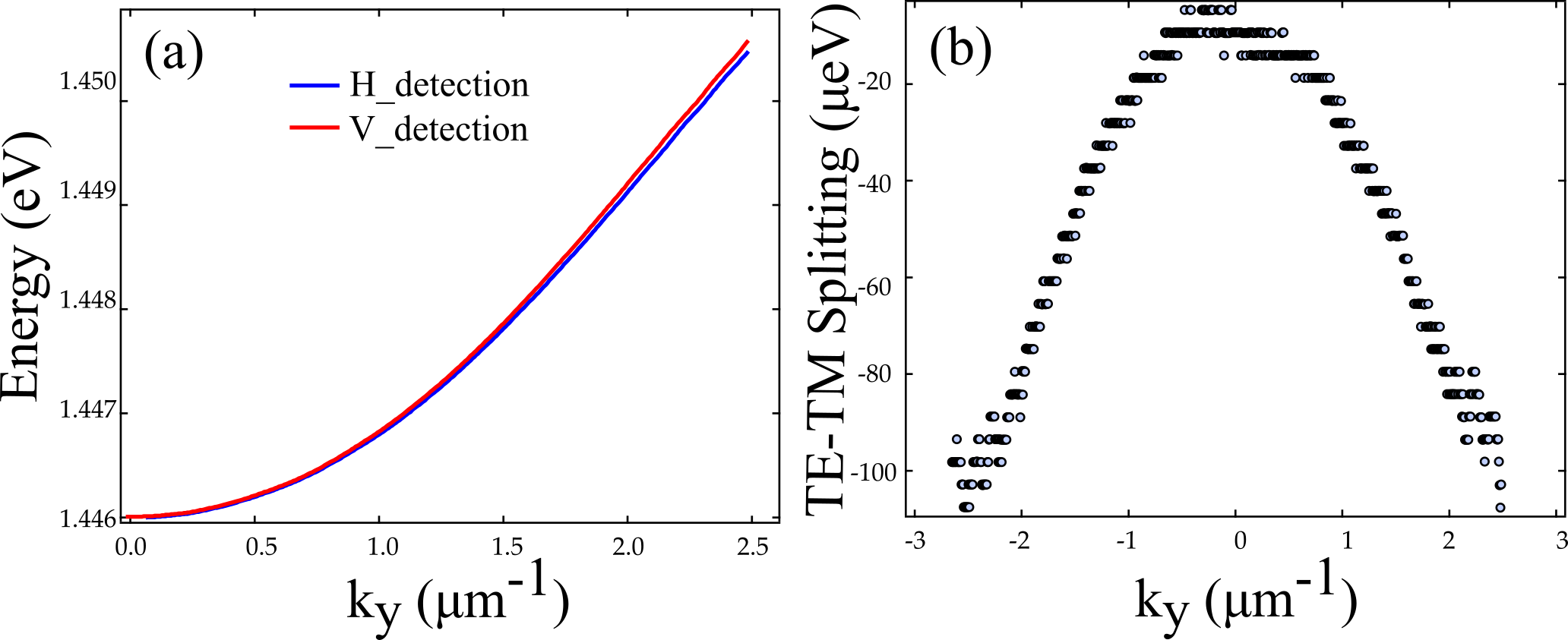}
    \caption{(a) Fitted lower polariton branch in horizontal (blue) and vertical (red) linear polarization detection. (b) Experimentally obtained TE-TM splitting for different wavevectors. }
    \label{tetm}
\end{figure}

We experimentally measure the TE-TM splitting of the sample by polarization resolving the lower polariton branch in the linear regime (i.e., below condensation threshold) along the $k_y$ momentum axis. We observe that vertically polarized polaritons possess higher energy than horizontally polarized polaritons [see Fig.~\ref{tetm}(a)]. The energy difference between these branches gives the TE-TM splitting which follows the expected parabolic trajectory [see Fig.~\ref{tetm}(b)].

\section{8-point excitation}
\begin{figure}[h]
    \centering
    \includegraphics[width=0.5\columnwidth]{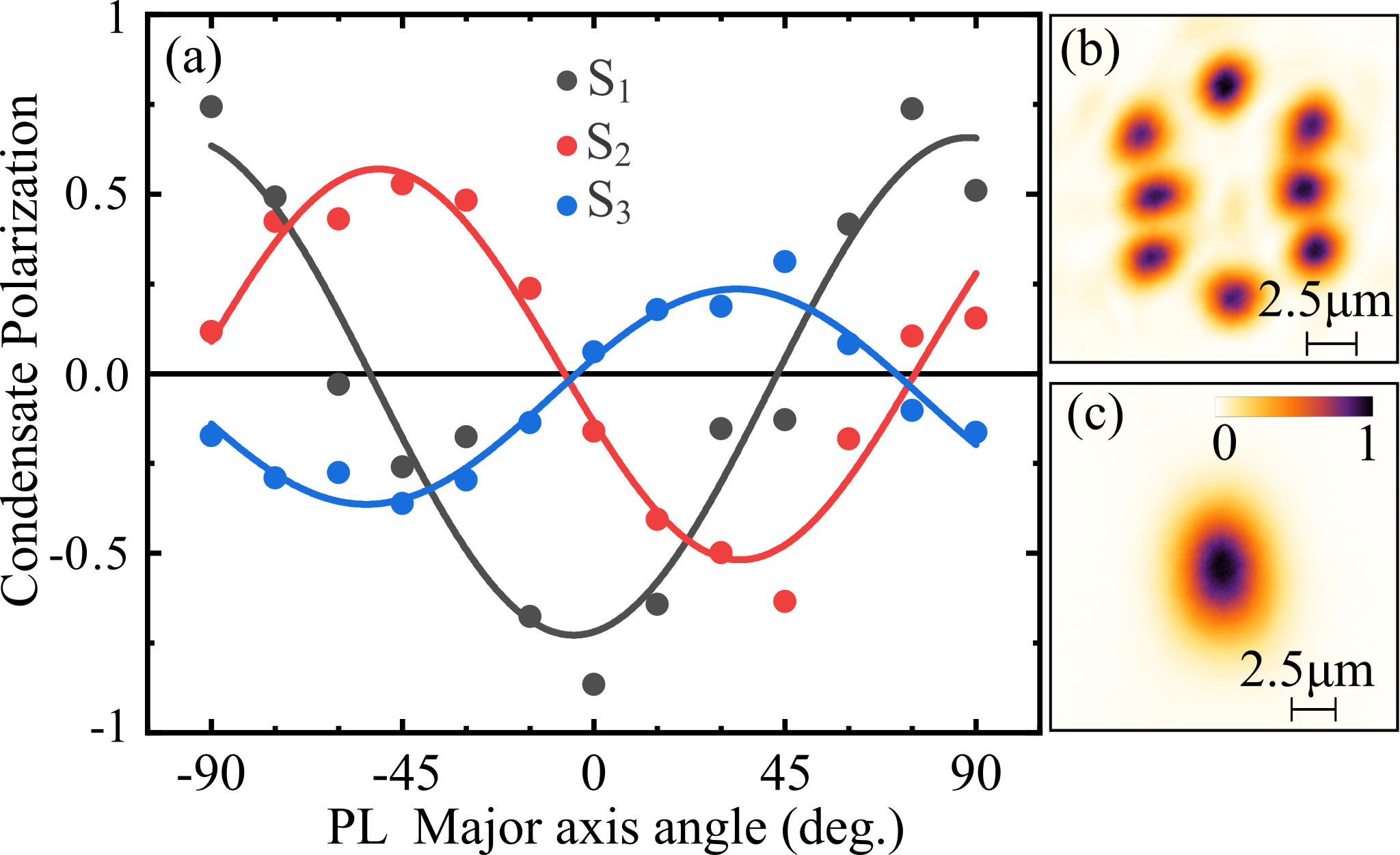}
    \caption{(a) Condensate polarization for different spatial orientations of the 8-Gaussian excitation profile major axis in real space at $P=2P_{th}$. (b) Excitation laser intensity profile. (c) Condensate PL.}
    \label{8point}
\end{figure}

In this supplemental section, we demonstrate the precise shape of our optical excitation beam is not important as long as it introduces different confinement strengths in the two orthogonal spatial directions. Here, we shape the overall laser profile using 8 Gaussians distributed in the form of an ellipse [see Fig.~\ref{8point}(b)] leading to the formation of an elliptical condensate [see Fig.~\ref{8point}(c)]. In agreement with the results presented in the main text, such an excitation profile also favors the formation of a condensate with linear polarization aligned along the ellipse minor axis. By rotating the excitation profile in the cavity plane Fig.~\ref{8point}(a), we observe the same rotation of the linear polarization of the condensate as in the main text. The deviations from the sinusoidal fits in Fig.~\ref{8point}(a) occur due to a some differences in power and shape of the individual Gaussian spots.

\section{The single-particle polariton Hamiltonian}
\begin{figure}
\centering
\includegraphics[width=0.8\linewidth]{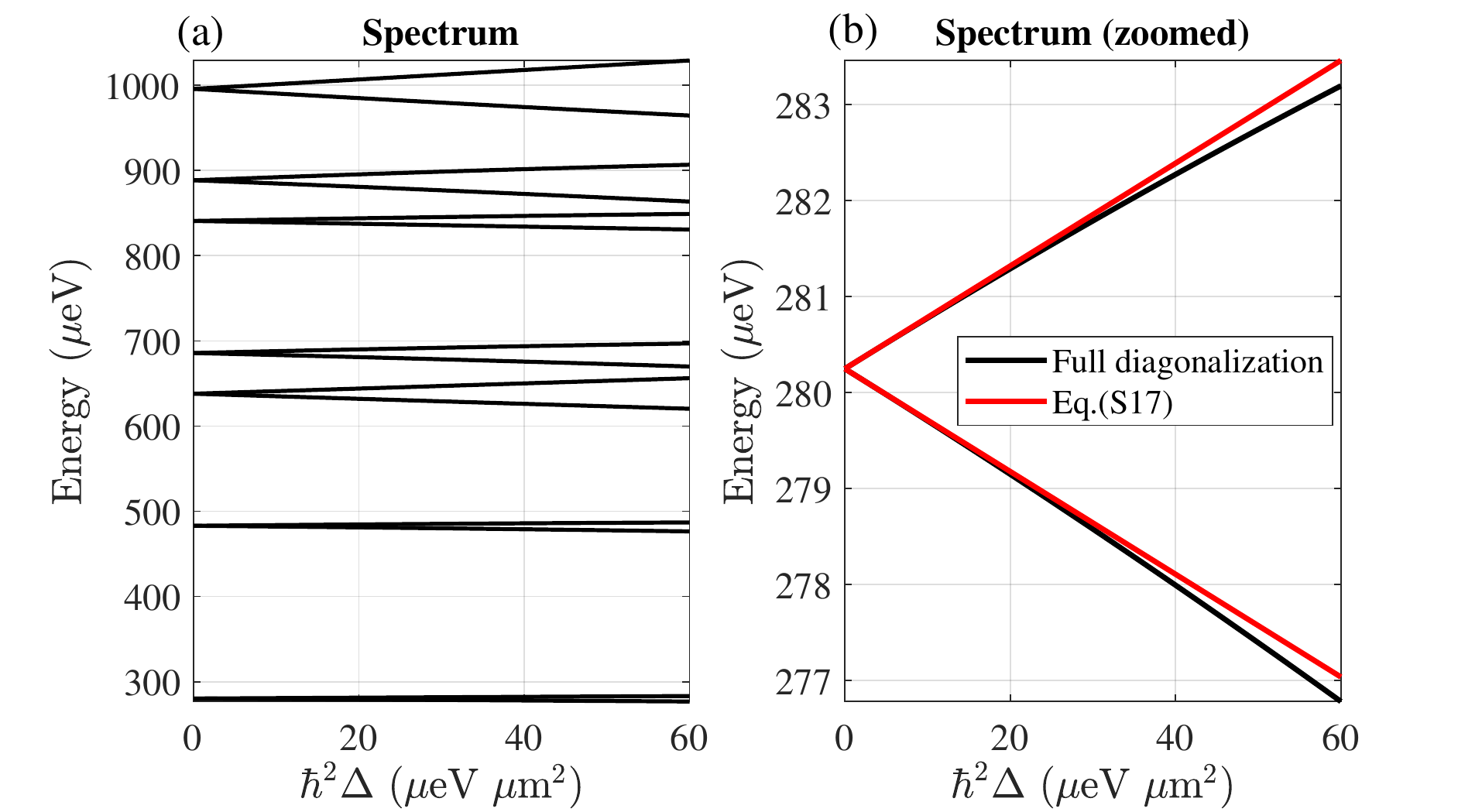}
\caption{Spectrum of the harmonic oscillator with TE-TM splitting and $\Gamma=0$ for simplicity. Black lines correspond to diagonalization of Eq.~\eqref{eq9} for $\text{max}{(n_{x(y)})} = 15$ and $\omega_x = 1.75 \omega_y =  0.55$ ps$^{-1}$ which were estimated from the full-width-half-maximum of the trapped condensate, and for a typical polariton mass of $m = 0.3$ meV ps$^2$ $\mu$m$^{-2}$. The red lines are outcome of Eq.~\eqref{eq.gs}.}
\label{fig1s}
\end{figure}

In the non-interacting (linear) regime the polaritons obey the following Hamiltonian (same as Eq.~(2) in the main text),
\begin{equation}
\hat{H} = \frac{\hbar^2 k^2}{2m} - \boldsymbol{\sigma} \cdot \boldsymbol{\Omega} + V(\mathbf{r}) - \frac{i \hbar \Gamma}{2},
 \end{equation}
where $m$ is the polariton mass, $\mathbf{k}=(k_x,k_y)$ is the in-plane cavity momentum, $\Gamma^{-1}$ is the polariton lifetime, $\boldsymbol{\sigma}$ is the Pauli matrix vector, and
\begin{equation}
 \boldsymbol{\Omega} = \hbar^2 \Delta (k_x^2 - k_y^2, \ 2 k_x k_y, \ 0)^T,
 \end{equation}
is the effective magnetic field [see Fig.~\ref{fig1s}(e)] coming from the TE-TM splitting of strength $\Delta$. We will consider that our laser generated potential in experiment can be approximated by an elliptically shaped harmonic oscillator (HO),
\begin{equation}
V(\mathbf{r}) = \frac{1}{2} m \omega_{x}^2 x^2 + \frac{1}{2} m \omega_{y}^2 y^2.
\end{equation}
Using the shorter momentum operator expression $p_{x(y)} = -i \hbar \partial_{x(y)} = \hbar k_{x(y)}$ for brevity, our Hamiltonian becomes:
\begin{equation} 
\hat{H} = \begin{pmatrix} \dfrac{p_x^2}{2m} +  \dfrac{p_y^2}{2m} +V(\mathbf{r}) -\dfrac{i \hbar \Gamma}{2}   & -\Delta (p_x - i p_y)^2 \\ 
-\Delta (p_x - i p_y)^{2\dagger}& \dfrac{p_x^2}{2m} +  \dfrac{p_y^2}{2m} + V(\mathbf{r})  -\dfrac{i \hbar \Gamma}{2} \end{pmatrix}.
\end{equation}
We will diagonalize this problem in the basis of the harmonic oscillator modes written for spin-up and spin-down particles as,
\begin{align}
|\psi_+ \rangle & = \sum_{n_x, n_y} c^{(+)}_{n_x, n_y} | n_x, n_y \rangle \\
|\psi_- \rangle & = \sum_{n_x, n_y} c^{(-)}_{n_x, n_y} | n_x, n_y \rangle, 
\end{align}
where $| n_x, n_y \rangle = | n_x \rangle \otimes | n_y \rangle$ are the harmonic oscillator eigenmodes in the ladder operator $\hat{a}_{x(y)}$ formalism. 
These are defined in the standard way through the position and momentum operators,
\begin{align}
x &= \sqrt{ \frac{\hbar}{2m \omega_{x}}} (\hat{a}^\dagger_{x} + \hat{a}_{x}), \qquad y = \sqrt{ \frac{\hbar}{2 m \omega_{y} }} (\hat{a}^\dagger_{y} + \hat{a}_{y}), \\
p_x &= i \sqrt{ \frac{m \hbar \omega_x}{2} } (\hat{a}_x^\dagger - \hat{a}_x), \qquad  p_y = i \sqrt{ \frac{m \hbar \omega_y}{2} } (\hat{a}_y^\dagger - \hat{a}_y).
\end{align}
Our Hamiltonian can then be expressed,
\begin{equation} \label{eq9}
\hat{H}= \begin{pmatrix} \hbar \omega_x \left(\hat{a}_x^\dagger \hat{a}_x + \dfrac{1}{2}\right) +  \hbar \omega_y \left(\hat{a}_y^\dagger \hat{a}_y + \dfrac{1}{2}\right) -\dfrac{i \hbar \Gamma}{2}  & -\Delta (p_x-i p_y)^2 \\ 
-\Delta (p_x - ip_y )^{2 \dagger} & \hbar \omega_x \left(\hat{a}_x^\dagger \hat{a}_x + \dfrac{1}{2}\right) +  \hbar \omega_y \left(\hat{a}_y^\dagger \hat{a}_y + \dfrac{1}{2}\right) -\dfrac{i \hbar \Gamma}{2}  
\end{pmatrix},
\end{equation}
where the following holds,
\begin{align}
p_{x(y)}^2 =(p_{x(y)}^2)^\dagger &= - \frac{m \hbar \omega_{x(y)}}{2} (\hat{a}_{x(y)}^\dagger \hat{a}_{x(y)}^\dagger  - \hat{a}_{x(y)}^\dagger \hat{a}_{x(y)} -   \hat{a}_{x(y)} \hat{a}_{x(y)}^\dagger +  \hat{a}_{x(y)} \hat{a}_{x(y)}), \\
p_x p_y  = (p_x p_y)^\dagger &= - \frac{m \hbar \sqrt{\omega_x \omega_y}}{2} (\hat{a}_x^\dagger \hat{a}_y^\dagger  - \hat{a}_x^\dagger \hat{a}_y -   \hat{a}_x \hat{a}_y^\dagger +  \hat{a}_x \hat{a}_y).
\end{align}
%
%
%
The diagonal harmonic oscillator terms can be written more neatly as,
\begin{equation}
E^{(0)}_{n_x,n_y}  = \hbar \omega_x \left( n_x + \frac{1}{2}\right) + \hbar \omega_y \left(n_y + \frac{1}{2}\right).
\end{equation}
The TE-TM terms will operate on our states as follows,
\begin{align} \notag
p_{x}^2 | n_x, n_y\rangle  = & - \frac{m \hbar \omega_{x}}{2} (\hat{a}_{x}^\dagger \hat{a}_{x}^\dagger  - \hat{a}_{x}^\dagger \hat{a}_{x} -   \hat{a}_{x} \hat{a}_{x}^\dagger +  \hat{a}_{x} \hat{a}_{x}) | n_x, n_y \rangle \\
 = & - \frac{m \hbar \omega_{x}}{2} (
\sqrt{n_x+1}\sqrt{n_x+2}| n_x+2, n_y \rangle  
- n_x | n_x, n_y \rangle 
- (n_x+1)| n_x, n_y \rangle 
+ \sqrt{n_x} \sqrt{n_x-1}| n_x-2, n_y \rangle), \\ \notag
p_x p_y | n_x, n_y\rangle  = &- \frac{m \hbar \sqrt{\omega_x \omega_y}}{2} (\hat{a}_x^\dagger \hat{a}_y^\dagger  - \hat{a}_x^\dagger \hat{a}_y -   \hat{a}_x \hat{a}_y^\dagger +  \hat{a}_x \hat{a}_y) | n_x, n_y \rangle \\ \notag
 =& - \frac{m \hbar \sqrt{\omega_x \omega_y}}{2}\bigg(
\sqrt{(n_x+1)(n_y + 1)} | n_x+1, n_y+1 \rangle 
- \sqrt{n_x+1}\sqrt{n_y} | n_x+1, n_y-1 \rangle \\
&-  \sqrt{n_x}\sqrt{n_y+1} | n_x-1, n_y+1 \rangle 
+  \sqrt{n_x n_y}| n_x-1, n_y-1 \rangle 
\bigg).
\end{align}
We will give a special notation to TE-TM terms which do not mix levels,
\begin{equation} \label{eq.S}
\epsilon_{n_x, n_y} = 
  -\frac{ m \hbar \Delta}{2}[\omega_{x}(2n_x + 1)  - \omega_{y}(2n_y + 1)].
\end{equation}
We can write a truncated version of our Hamiltonian for just the spins in the trap ground state $|0,0\rangle$ which reads (i.e., coupling to other HO levels is neglected), 
\begin{equation} \label{eq.Htrun}
\hat{H} \approx \begin{pmatrix}
 E^{(0)}_{0,0} -\dfrac{i \hbar \Gamma}{2} & \epsilon_{0,0}  \\
\epsilon_{0,0} &  E^{(0)}_{0,0} -\dfrac{i \hbar \Gamma}{2} 
\end{pmatrix}.
\end{equation}
The eigenvectors are the horizontally (H) and vertically (V) polarized states of light with eigenvalues,
\begin{equation}\label{eq.gs}
E_{0,0}^{(H,V)} = \frac{\hbar}{2} \left[ \omega_x + \omega_y \mp  m \Delta(\omega_{x}  - \omega_{y}) \right] -\frac{i \hbar \Gamma}{2}.
\end{equation}
We remind that $\Delta<0$ in our cavity sample~\cite{Maragkou_OptLett2011s} (see Fig.~\ref{tetm}). This expression confirms experimental observations of the cavity energy resolved emission in Fig.~4(b) in the main text. When the laser induced trap has a major axis along the vertical direction (i.e., $\omega_x > \omega_y$) then we observe higher frequency in the horizontally emitted light as opposed to the vertical light, in agreement with $E_{0,0}^{(H)}>E_{0,0}^{(V)}$. When $\omega_x < \omega_y$ the vice versa appears. In Fig.~\ref{fig1s} we put $\Gamma=0$ for simplicity and compare the calculated spectrum obtained from diagonalizing Eq.~\eqref{eq9} (black lines) for $\text{max}{(n_{x(y)})} = 15$ modes against our truncated lowest HO level Hamiltonian Eq.~\eqref{eq.gs}. The generalization of Eq.~\eqref{eq.Htrun} to arbitrary angles of the potential orientation in the $x$-$y$ plane is straightforward and presented in Eq.~(4) in the main text.

\section{Generalized Gross-Pitaevskii simulations}
We will now model the dynamics of the polariton condensate spinor $\Psi = (\psi_+, \psi_-)^T$ using the generalized (driven-dissipative) Gross-Pitaevskii equation, coupled to a semiclassical rate equation describing a reservoir of low-momentum excitons $\mathbf{X} = (X_+,X_-)^T$ which scatter into the condensate~\cite{Wouters2007as}:
\begin{equation}\label{gpe}
\begin{split}
i\hbar {\partial\psi_\pm \over\partial t} & = \left[ - { \hbar^2 \nabla^2 \over 2 m} - \Delta (p_x \mp i p_y)^2 + g\left(X_\pm + \frac{P}{W}\right) +  \alpha \vert \psi_\pm \vert^2 + i \hbar {RX_\pm - \Gamma \over 2}  \right] \psi_\pm,\\
{\partial X_\pm \over \partial t} & = P - (R \vert \psi_\pm \vert^2 + \Gamma_R) X_\pm + \Gamma_s(X_{\mp} - X_\pm). 
\end{split}
\end{equation}
Here, $\nabla^2$ denotes the two-dimensional Laplacian operator, $\Delta$ the TE-TM splitting, $m$ is the  polariton effective mass, $g$ and $\alpha$ are the interaction constants describing the polariton repulsion off the exciton density and polariton-polariton repulsion, $R$ governs the stimulated scattering from the reservoir into the condensate, $\Gamma$ and $\Gamma_R$ are the polariton and active exciton decay rates, and $\Gamma_s$ is the rate of spin relaxation. As we are working in continuous wave regime, and the pump is linearly polarized at all times, we do not need to take into account the polarization- and time-dependence of a high-momentum (inactive) reservoir describing excitons that are too energetic to scatter into the condensate~\cite{Anton_PRB2013s}. Instead, the contribution of photoexcited high-momentum excitons to the condensate appears through the blueshift term $P/W$ where $W$ describes the conversion rate of high-momentum excitons into low-momentum excitons $X_\pm$ that sustain the condensate.

We will use the pseudospin formalism (analogous to the Stokes parameters describing the cavity photons) to describe the polarization of the condensate (note that in Eq.~(1) in the main manuscript we have used the normalized definition),
\begin{equation} \label{eq.pseudospin}
    \mathbf{S} = \begin{pmatrix}
    S_1 \\
    S_2 \\
    S_3 
    \end{pmatrix} = \Psi^\dagger \boldsymbol{\sigma} \Psi.
\end{equation}
Here, $\boldsymbol{\sigma}$ is the Pauli matrix vector and the total density of the condensate is expressed as $S_0 =  |\psi_+|^2 + |\psi_-|^2$. 

We will study the dynamics of Eq.~\eqref{gpe} for three different excitation profiles shown in Figs.~\ref{figs2}(a,d,f). The profiles shown in Fig.~\ref{figs2}(d) and~\ref{figs2}(f) represent the experimental configurations shown in Fig.~1(a,b) in the main manuscript and in Fig.~\ref{8point}(b). We also introduce, for completeness, a third type of an elliptical excitation profile in the numerical analysis shown in Fig.~\ref{figs2}(a). The three excitation profiles can be written as follows,
\begin{align} \label{eq.pump1}
    P_I(\mathbf{r}) & = P_0 \frac{L_I^4}{\left(\dfrac{x^2}{a^2} + \dfrac{y^2}{b^2} - r_0^2\right)^2 + L_I^4}, \\ \label{eq.pump2}
       P_{II}(\mathbf{r}) & = P_0 \frac{L_{II}^4}{\left(x^2 + y^2 - r_0^2\right)^2 + L_{II}^4} [1 - \eta \cos{(2 \varphi + \phi_\text{maj})} \sin^2{(\pi r / 2 r_0)}], \\ \label{eq.pump3}
   P_{III}(\mathbf{r}) & = P_0 \sum_{n=1}^8 \frac{L_{III}^2}{(x-x_n)^2 + (y-y_n)^2 + L_{III}^2}.
\end{align}
Here, $L_{I,II,III}$ denote the spread (thickness) of the potentials. For pumps~\eqref{eq.pump1} and~\eqref{eq.pump2} the common radius $r_0$ defines the length of the ellipse minor and major axis. Specifically, the minor and major axis are given by the parameters $0<a,b<1$ for ~\eqref{eq.pump1} and $\eta,\phi_\text{maj}$ for~~\eqref{eq.pump2}. For the third pump profile~\eqref{eq.pump3} we use coordinates $x_n,y_n$ of the eight tightly focused pump spots corresponding to the experiment. The laser power density is given by $P_0$.

\begin{figure}[t]
\centering
\includegraphics[width=\linewidth]{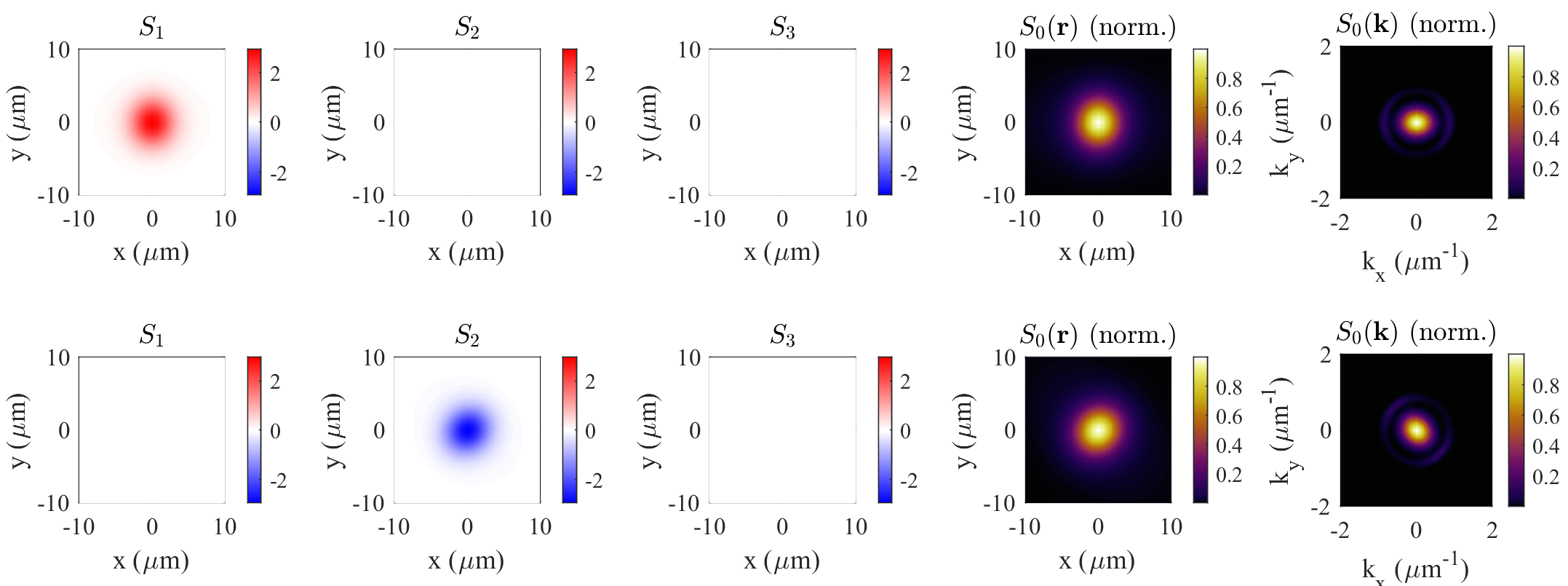}
\caption{(Top row) Steady state solution from simulation of Eq.~\eqref{gpe}. Here we used the $P_I(\mathbf{r})$ pump profile with a major axis along the vertical direction, $b/a = 1.3$. The results show a condensate forming with horizontal polarization implying occupation of the excited trap spin state. (Bottom row) Same simulations but now the pump major axis is orientated $\pi/4$ from the horizontal resulting in antidiagonally polarized condensate which again corresponds to the excited spin state of the trap.}
\label{fig1_2}
\end{figure}

\begin{figure}
\centering
\includegraphics[width=0.9\linewidth]{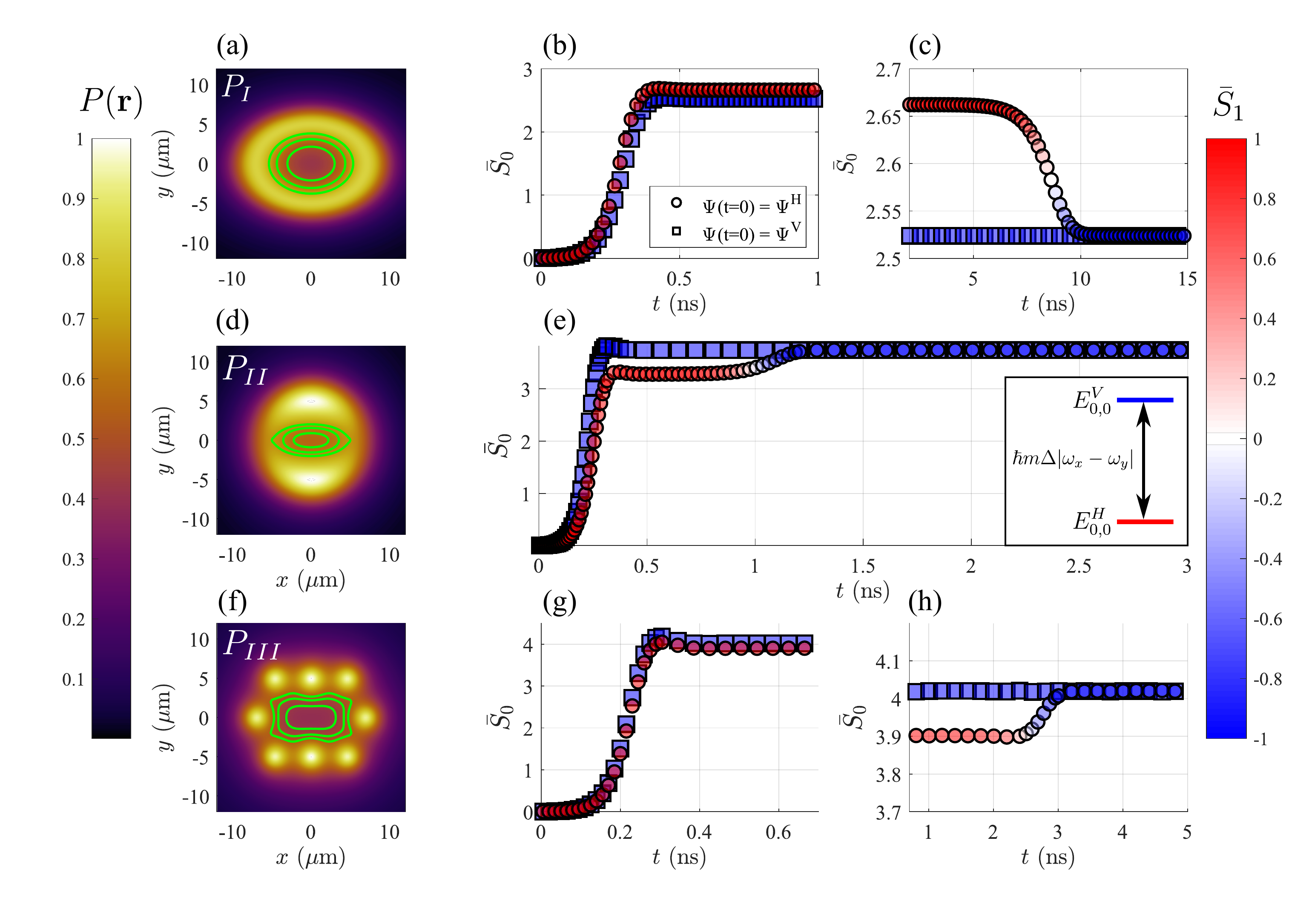}
\caption{(a,d,f) Pump profiles (normalized) corresponding to Eqs.~\eqref{eq.pump1}-\eqref{eq.pump3}. In (a) we have set $a/b = 1.4$ and in (d) $\phi_\text{maj}=0$ and $\eta =0.2$. The green lines are contours which represent the isoenergy lines of the optical trap. The inset in (e) shows schematically the energy structure of the linearly polarized modes in the trap s-orbital. In the panels on the right we show the corresponding condensate $\bar{S}_0$ dynamics for the two ansatz $\Psi^{H,V}$ in Eq.~\eqref{eq.ansatz} separately (circles and squares respectively) for fixed pumping value $P_0$ above threshold. The red-blue color denotes the area-integrated linear polarization $\bar{S}_1$. In panels (b,c) and (g,h) we split the time axis to better show the early and late dynamics. In panel (b) the early dynamics show that the $\Psi^H$ ansatz rises faster and the condensate saturates into a horizontally polarized state (red circles) with higher particle number as compared to the vertically polarized state (blue squares), implying it has larger gain. At later times shown in panel (c), the horizontal polarized solution always collapses into the vertically polarized solution, indicating that it is only metastable and that a vertically polarized condensate survives at long times. In panels (e) and (g) we observe a qualitatively different early dynamics where now the horizontally polarized condensate rises slower. Eventually, as we see in (e) and (h), the horizontally polarized solution destabilizes and converges into a vertically polarized condensate.}
\label{figs2}
\end{figure}

In Fig.~\ref{fig1_2} we show the obtained steady state wavefunction $\Psi$ obtained from random initial conditions while driving the system above the pump threshold $P_0>P_\text{th}$ using the first pump profile $P_I(\mathbf{r})$. The threshold $P_\text{th}$ is defined as the transition point where the normal state $|\Psi| = 0$ becomes unstable and instead a condensate forms $|\Psi|>0$. Indeed, choosing parameters corresponding to the experiment we obtain complete match between experimental observations and simulations. Testing 100 different random initial conditions we find that the simulated condensate always converges to a steady state corresponding to the excited spin state of the trap. The parameters of the simulation are: $m = 0.3$ meV ps$^2$ $\mu$m$^{-2}$; $\Gamma = \frac{1}{5.5}$ ps$^{-1}$; $\hbar \alpha = 3$ $\mu$eV  $\mu$m$^2$, $g = \alpha$; $\Gamma_R = \Gamma/4$; $R = 0.67 \alpha$; $W = 0.05$ ps$^{-1}$; $\Gamma_s = \Gamma_R/2$; $\hbar^2 \Delta = -0.03$ meV $\mu$m$^2$; $L_I = 5$ $\mu$m; $L_{II} = 6$ $\mu$m; $L_{III} = 2$ $\mu$m; $r_0 = 5$ $\mu$m; $\eta = 0.2$; and $P_0 = 4.625$ $\mu$m$^{-2}$ ps$^{-1}$ which is around $\approx 20\%$ above threshold for each configuration.

\subsection{Condensate metastability}
We will here scrutinize the early and late condensate dynamics using two simple initial conditions (ansatz). We will use a trap with a horizontal major axis (i.e., $\omega_x<\omega_y$) since all other major axis orientations are completely generalizable. The two initial conditions for Eq.~\eqref{gpe} are written,
\begin{equation}\label{eq.ansatz}
\Psi(t=0) = \Psi^{H,V} = A e^{-c (x^2+y^2)} \begin{pmatrix}
1 \\ \pm 1
\end{pmatrix}.
\end{equation}
The parameter $c$ is chosen to have $\Psi^{H,V}$ localized dominantly within the trap and $A\ll1$ is a small number to minimize nonlinear effects in the initial dynamics. We then solve Eq.~\eqref{gpe} for each initial condition and plot the spatially integrated particle number $S_0$ and $S_1$ (normalized) Stokes parameter as a function of time,
\begin{equation}
\bar{S}_0(t) = \frac{1}{\mathcal{A}}\int_\mathcal{A} S_0(\mathbf{r},t) d\mathbf{r} \qquad \bar{S}_1(t) = \frac{ \int_\mathcal{A} S_1(\mathbf{r},t) d\mathbf{r}}{ \int_\mathcal{A} S_0(\mathbf{r},t) d\mathbf{r}},
\end{equation}
where $\mathcal{A}$ is the area enclosed by the pump profile ridge (i.e., $\text{max}\,{P(\mathbf{r})}$).

Simulations using the generalised Gross-Pitaevskii equation~\eqref{gpe} for fixed power $P_0$ (above threshold) and the three pump profiles $P_{I,II,III}$ are shown in Figs.~\ref{figs2}(b,c), and~\ref{figs2}(e), and~\ref{figs2}(g,h), respectively. We plot the area-integrated particle number $\bar{S}_0$ for the two different initial conditions $\Psi^{H,V}$ (circles and squares, respectively). The color of the markers indicates the area-integrated linear polarization $\bar{S}_1$ of the condensate. In Figs.~\ref{figs2}(b,c) and~\ref{figs2}(g,h) we split the time axis to show better the early and late dynamics. The inset in Fig.~\ref{figs2}(e) shows schematically the energy splitting between the polarizations. For pump profiles $P_{II}$ and $P_{III}$ we see that in the early dynamics a vertically polarized condensate (blue squares) rises faster and saturates at a higher particle number than a horizontally polarized condensate (red circles). This can be understood from the fact that the vertical and horizontal polarized modes of the condensate have different effective masses and, thus, have different penetration depths into the gain region of the pump. In particular, pumps $P_{II}$ and $P_{III}$ lead to an excess density of reservoir excitons about the short axis of the ellipse. Since the penetration depth of the confined mode in the potential well is larger in the direction of the linear polarization axis, this would increase the overlap of the mode co-polarized with the short-axis of the potential well with the gain region, i.e. the 'excited state' in the fine structure. In the late dynamics [Fig.~\ref{figs2}(e) and~\ref{figs2}(h)] the horizontal solution destabilizes and converges into the vertically polarized solution. This is in agreement with our experimental observations showing robust condensation into the excited spin state of the optical traps.

Interestingly, for pump $P_I$ the early dynamics [Fig.~\ref{figs2}(b)] are reversed with respect to $P_{II,III}$. Now the horizontally polarized condensate rises faster and saturates at a higher particle number. This pump profile does not generate a strong excess of excitons about the ellipse short axis and, as a consequence, the higher energy polaritons, co-polarized with the short axis, escape (leak) faster from the trap. Nevertheless, in the late dynamics [Fig.~\ref{figs2}(c)] the horizontally polarized solution destabilizes and collapses into the vertically polarized solution. This observation underlines that the stable solution of the condensate does not necessarily correspond to the one with maximum particle number $\bar{S}_0$. In the next section we address the different parameters of our model that determine the stability of the excited state and the ground state.

\section{Stability analysis on a two mode problem}
Here, we will analyse the stability properties of the condensate by projecting our order parameter on only the lowest trap state $|0,0\rangle$. Let us here denote $\psi_\pm(\mathbf{r},t) \to \psi_\pm(t)$ as the condensate order parameter describing polaritons only in the HO ground state and neglect contribution from higher HO modes,
\begin{align} \label{eq.orig}
\begin{split}
i & \frac{ d\psi_\pm}{dt}  = \Big[ \alpha |\psi_\pm|^2 + g X_\pm + i\frac{R X_\pm  - \Gamma}{2}  \Big] \psi_\pm + (\epsilon + i \gamma) \psi_{\mp}, \\ 
& \frac{ dX_\pm}{dt}  =   P - \left( \Gamma_R + R |\psi_\pm|^2 \right)X_\pm + \Gamma_s(X_{\mp} - X_\pm).
\end{split}
\end{align}
The parameters here have the same meaning as their counterparts in Eq.~\eqref{gpe}, but we stress that we have absorbed $\hbar$ into their definition for brevity and some will obtain modified values after integrating out the spatial degrees of freedom depending on the precise shape of the condensate and reservoir. We have removed the $P/W$ term as it only induces an overall blueshift to both spins which does not affect the stability properties of the system. The spin-coupling parameter $\epsilon$ corresponds to $\epsilon_{0,0}$ from Eq.~\eqref{eq.S}. We additionally include an imaginary coupling parameter $\gamma$ which physically represents different linewidths of the horizontal and vertical polarized modes (i.e., different decay rates).

The two steady state solutions of interest correspond to spin-balanced reservoirs $X_+ = X_-$ supporting either purely horizontally or vertically polarized condensate written,
\begin{equation}
\Psi_\text{st}^{H,V} = \sqrt{\frac{S_0^{H,V}}{2}} 
\begin{pmatrix} 
1 \\ 
\pm 1
 \end{pmatrix} e^{-i \omega^{H,V} t}
\end{equation}
where,
\begin{equation}
\omega^{H,V} = \frac{\alpha}{2} S_0^{H,V}  + g X^{H,V} \pm \epsilon_{0,0}, \qquad S_0^{H,V} = \frac{P}{\Gamma \mp 2\gamma} - \frac{\Gamma_R}{R}, \qquad X^{H,V} = \frac{\Gamma \mp 2\gamma}{R}.
\end{equation}
The power to reach the lower threshold solution is $P_\text{th} = \Gamma_R(\Gamma - 2 |\gamma|)/R$. The stability analysis of the $\Psi_\text{st}^{H,V}$ solutions is exactly the same as in Ref.~\cite{Sigurdsson_PRR2020s} where a $5\times5$ Jacobian matrix $\mathbf{J}$ corresponding to linearisation of Eq.~\eqref{eq.orig} around its steady state solutions was derived. This allows determining the stability of the solutions in terms of their Jacobian eigenvalues $\lambda_n$ (also known as {\it Lyapunov exponents} in nonlinear dynamics or {\it Bogoliubov elementary excitations} in the context of Bose-Einstein condensates). If a single eigenvalue of $\mathbf{J}$ has a positive real part then the solution is said to be {\it asymptotically unstable}. It was found in Ref.~\cite{Sigurdsson_PRR2020s} that the relative strength between the  mean field energy coming from the condensates $\alpha |\psi_\pm|^2$ and the reservoir blueshift $g X_\pm$ played a big role in whether the excited state or the ground state was stable. 

To understand this better, we will first consider the stability of the two solutions $\Psi_\text{st}^{H,V}$ using a more general coupled Gross-Pitaevskii equations (similar to coupled amplitude oscillators),
\begin{equation}\label{eq.simp}
i \frac{d\psi_\pm}{dt} = \alpha |\psi_\pm|^2 \psi_\pm + \epsilon \psi_\mp, \qquad \epsilon>0.
\end{equation}
Clearly, $\Psi_\text{st}^{H,V} $ is a solution of the above equation for any particle number $S_0 =S_0^H = S_0^V$ with frequency $\omega^{H,V} = \alpha S_0/2 \pm \epsilon$. The Lyapunov exponents of these solutions are written:
\begin{equation}\label{lyp}
\begin{split}
\lambda_1^{H,V} & = 0, \\
\lambda_{2}^{H,V} &= \,  \, \,  \, \sqrt{2\epsilon(-2\epsilon \pm \alpha S_0)},\\
\lambda_{3}^{H,V} &= - \sqrt{2\epsilon(-2\epsilon \pm \alpha S_0)}.
\end{split}
\end{equation}
We only have three eigenvalues because our two level system~\eqref{eq.simp} can be described with the three-dimensional pseudospin state vector [see Eq.~\eqref{eq.pseudospin}], analogous to the Stokes vector of light. It is clear that only $\lambda_{2}^{H,V}$ can have real values greater than zero (the signature of instability) and there are only two cases when this happens:
\begin{enumerate}
\item If $\alpha>0$ (repulsive particle interactions) then $\text{Re}{(\lambda_{2}^{\color{red} H})} > 0$ when $\alpha S_0 > 2\epsilon$.
\item If $\alpha<0$ (attractive particle interactions) then $\text{Re}{(\lambda_{2}^{\color{blue} V})} > 0$ when $|\alpha S_0| > 2\epsilon$.
\end{enumerate}
This simple result shows that the stability of the excited state and the ground state changes when the mean field energy $\alpha S_0$ exceeds the fine structure splitting $2 \epsilon$. In our case, polariton interactions are repulsive $\alpha>0$ and the condensate should always form in the ground state when $S_0>2\epsilon/\alpha$~\cite{Shelykh_PRL2006}. Note that $2 \hbar \epsilon \approx 20$ $\mu$eV in our experiment which is small compared to the typical polariton mean field energies $\alpha S_0$. It therefore appears puzzling that we observe stable excited state condensation when the above simple consideration dictates that only the ground state should be stable.  

\begin{figure}[b]
\centering
\includegraphics[width=0.8\linewidth]{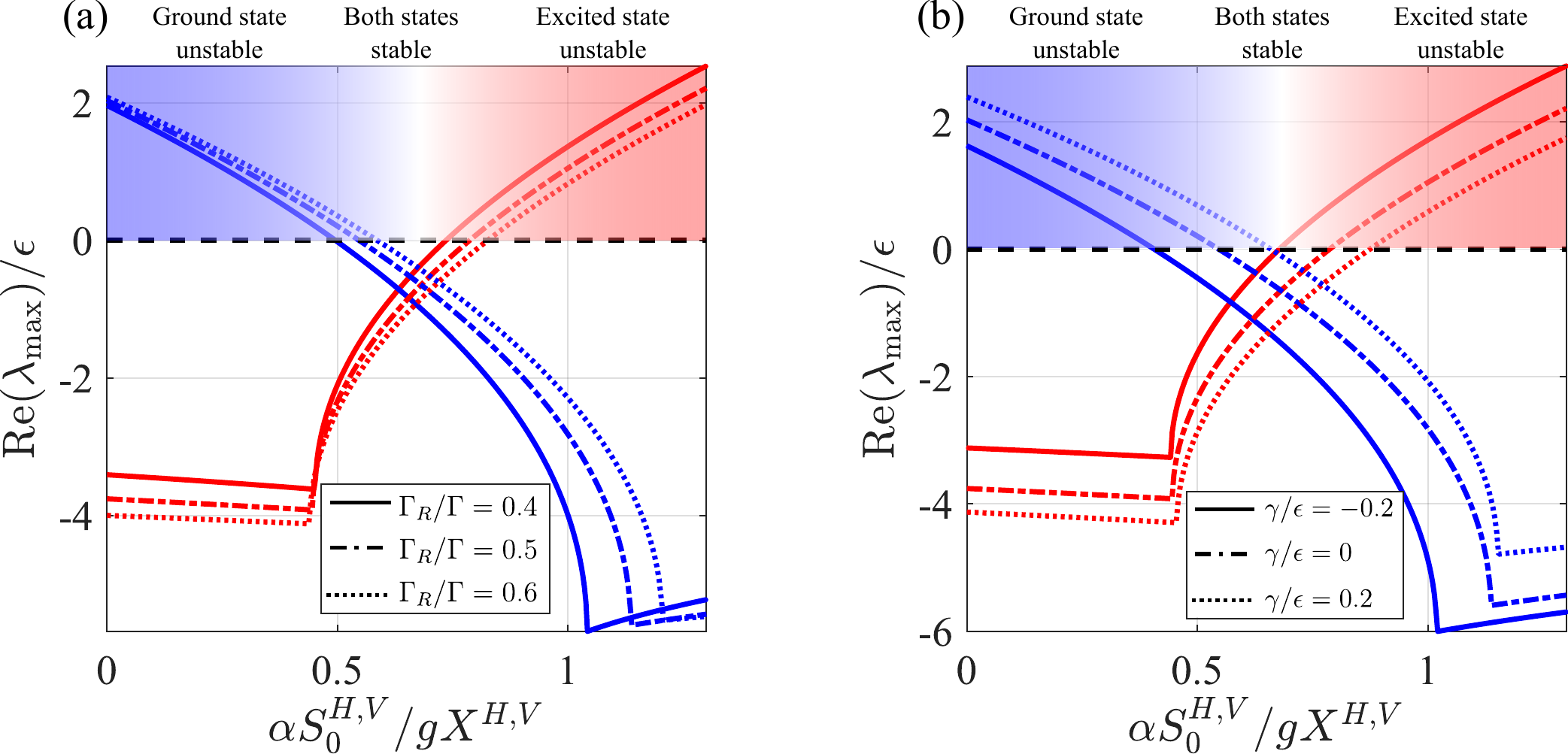}
\caption{Maximum real eigenvalues (Lyapunov exponents) from the Jacobian of Eq.~\eqref{eq.orig} around the two steady state solution $\Psi_\text{st}^{H,V}$ (red and blue curves, respectively) for $\epsilon=0.01$ ps$^{-1}$. Regimes where $\text{Re}{(\lambda_\text{max})}>0$ imply instability of the solution. The color gradient at the top is added for illustration purposes. Parameters: $\Gamma = 1/5$ ps$^{-1}$, $\Gamma_s = \Gamma/4$, $\epsilon = \Gamma/20$, $R = 0.015\Gamma$, $g = 5R/6$, and $P = 2P_\text{th}$. In (a) we fix $\gamma=0$ and in (b) we fix $\Gamma_R = 0.5 \Gamma$.}
\label{figs3}
\end{figure}

In the following, we will address two different mechanisms that fight against ground state condensation. \textbf{\underline{First,}} if the ground state is lossier than the excited state (i.e., $\gamma/\epsilon < 0$) then polaritons will preferentially condense into the excited state until $S_0$ exceeds a critical value~\cite{Sigurdsson_PRR2020s}. However, as we can see from Fig.~\ref{figs2}(b,c), even if the excited state is lossier than the ground state the condensate can still preferentially populate and stabilize in the excited state. \textbf{\underline{Second,}} a stable excited spin state condensation can appear due to an effective attractive nonlinearity coming from the reservoir [i.e., the term $g X_\pm$ in Eq.~\eqref{eq.orig}]. Indeed, it is well established that the presence of the condensate ``eats away'' the reservoir density analogous to the hole burning effect in lasers~\cite{Estrecho_NatComm2018s}. In the adiabatic regime where the reservoir is assumed to adjust to the condensate density dynamics very fast it can be approximated as follows,
\begin{equation}
X_\pm = \frac{P}{\Gamma_R} \left[ 1 - \frac{R |\psi_\pm|^2}{\Gamma_R} + \mathcal{O}(|\psi_\pm|^4) \right].
\end{equation}
The nonlinearity of the condensate can therefore described by an effective interaction parameter,
\begin{equation} \label{eq.alpha_eff}
\alpha_\text{eff} = \alpha - \frac{g P R}{\Gamma_R^2}.
\end{equation}
To test our hypothesis, we numerically solve the eigenvalues of the Jacobian for Eq.~\eqref{eq.orig} and plot their maximum real part for the $\Psi_\text{st}^{H,V}$ solutions in Fig.~\ref{figs3} (red and blue curves, respectively) as a function of varying $\alpha$ and several values of $\Gamma_R$ [Fig.~\ref{figs3}(a)] and $\gamma$ [Fig.~\ref{figs3}(b)]. Indeed, we see that there exist three distinct regimes which we schematically illustrate with the blue-white-red color gradient:
  \begin{center}
    \begin{minipage}{0.5\textwidth}
      \begin{enumerate}
    \item[i.] Ground state unstable and excited state stable.
    \item[ii.] Both states stable.
    \item[iii.] Ground state stable and excited state unstable.
      \end{enumerate}
    \end{minipage}
  \end{center}
As expected from Eq.~\eqref{eq.alpha_eff}, the stability range of the excited state increases as $\Gamma_R$ decreases [see Fig.~\ref{figs3}(a)] due to the nonlinearity $\alpha_\text{eff}$ becoming more negative. Moreover, the stability range of the excited state also increases when $\gamma/\epsilon$ becomes more negative corresponding to the ground state becoming lossy [see Fig.~\ref{figs3}(b)]. We point out that it is not possible to determine separately the contribution of $\alpha S_0^{H,V}$ and $g X^{H,V}$ in experiment since we can only measure the net blueshift in condensate energy. Nevertheless, our experimental results indicate that the current pump configuration favours the far-left regime in Fig.~\ref{figs3} where the ground state is unstable. Recently, we reported results corresponding to the far-right regime where robust ground state condensation was instead observed~\cite{Gnusov_PRB2020s} using the same cavity sample but somewhat different experimental configuration. How exactly one can tune from one regime to the other is difficult to tell, but the clearest path would either involve changing the detuning between the photon and exciton mode. This is possible because $\alpha \propto |\chi|^4$ and $g \propto |\chi|^2$ where $|\chi|^2$ is the exciton Hopfield fraction of the polariton quasiparticle. Another method would be to design an excitation profile $P(\mathbf{r})$ which changes the mean field rate $R$ of particles scattering into the condensate.

\begin{figure}
\centering
\includegraphics[width=\linewidth]{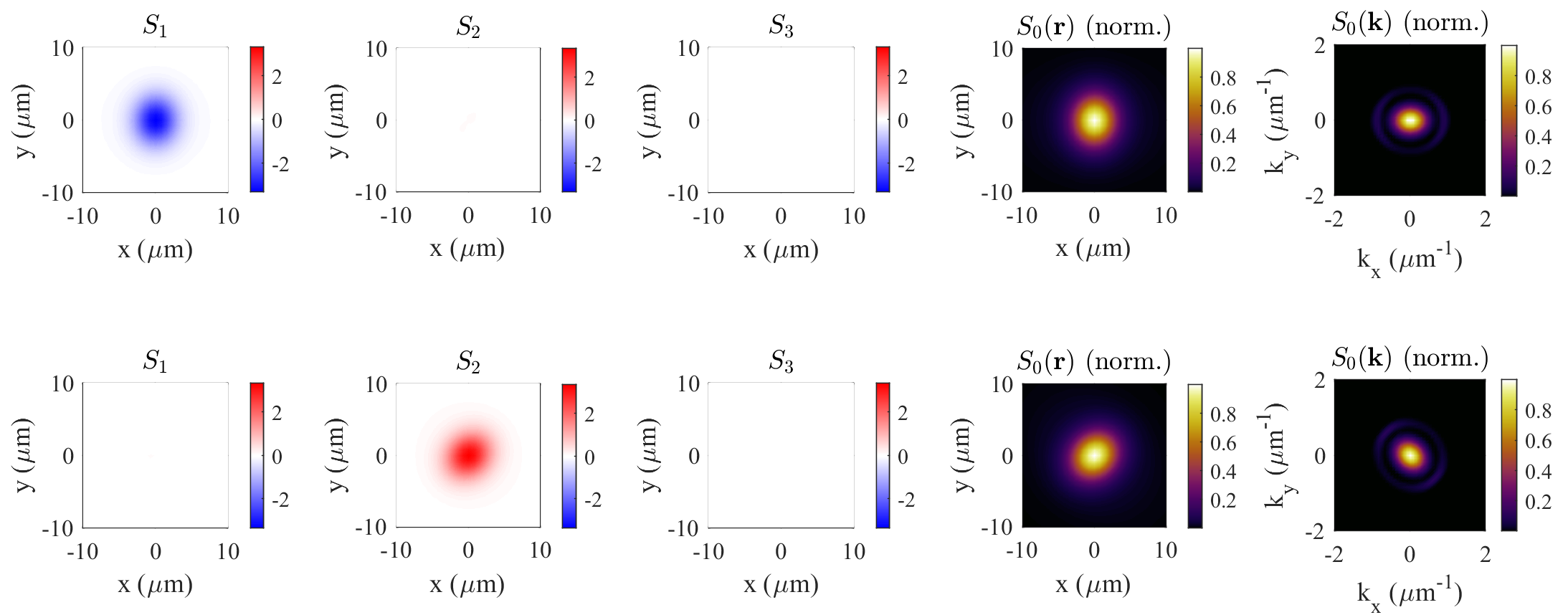}
\caption{Same simulations as in Fig.~\ref{fig1_2} but with doubled interactions strength $\alpha \to 2 \alpha$ (keeping all other parameters unchanged) such that the excited spin state now becomes unstable and only the ground state becomes populated.}
\label{figs5}
\end{figure}

Finally, to see if our hypothesis agrees with the full spatial calculations of Eq.~\eqref{gpe} we repeat the simulation from Fig.~\ref{fig1_2} in a new Fig.~\ref{figs5} with the strength of polariton-polariton interactions doubled, i.e. $\alpha \to 2 \alpha$ while keeping all other parameters unchanged. We now find, in agreement with our predictions, that the steady state solution (tested over 100 random initial conditions) converges to the spin ground state instead of the excited state. This can be evidenced from the opposite polarization appearing in the Stokes components in Fig.~\ref{figs5} as compared to Fig.~\ref{fig1_2}.

\section{Modeling of coupled optically trapped spinor condensates}
In this section we modify Eq.~\eqref{eq.orig} to describe coupling between two spatially separated condensates as shown in Fig.~5 in the main manuscript. The inter-condensate-coupling is sometimes referred to as {\it ballistic} coupling because energetic polaritons escape from each trap and undergo finite-time free-space propagation before they reach their neighboring condensate. Such coupling is qualitatively different from evanescent coupling (e.g., tunneling between Bose-Einstein condensates) since the propagation time of polaritons between the condensates is comparable to their intrinsic frequencies. This implies that the coupling between ballistic condensates is {\it time delayed}~\cite{Topfer_CommPhys2020} which we can introduce explicitly to Eq.~\eqref{eq.orig},
\begin{align} \label{eq.orig2}
\begin{split}
i & \frac{ d\psi_\pm^{(1,2)}}{dt}  = \Big[\omega_0 + \alpha |\psi_\pm^{(1,2)}|^2 + g X_\pm^{(1,2)} + i\frac{R X_\pm^{(1,2)}  - \Gamma}{2}  \Big] \psi_\pm^{(1,2)} + (\epsilon + i \gamma) \psi_{\mp}^{(1,2)} + J \psi_\pm^{(2,1)}(t - \tau) +  \mathcal{J} \psi_\mp^{(2,1)}(t - \tau), \\ 
& \frac{ dX_\pm^{(1,2)}}{dt}  =   P - \left( \Gamma_R + R |\psi_\pm^{(1,2)}|^2 \right)X_\pm^{(1,2)} + \Gamma_s(X_{\mp}^{(1,2)} - X_\pm^{(1,2)}).
\end{split}
\end{align}
Here, the indices $(1,2)$ refer to the two different condensates. We have also introduced the condensate intrinsic energy $\omega_0$ since a suitable rotating reference frame cannot be chosen for time delayed coupling between oscillators. As was previously demonstrated~\cite{Topfer_CommPhys2020}, the strength of the coupling $J$ depends on the separation distance $d$ between the condensates,
\begin{equation} \label{eq.hank}
    J(d) = J_0 |H_0^{(1)}(k_c d)|,
\end{equation}
where $H_0^{(1)}$ is the zeroth order Hankel function, $J_0 \in \mathbb{C}$ quantifies the non-Hermitian coupling strength dictated by the overlap of the condensates over the optical trap region, and $k_c$ is the complex wavevector of the polaritons propagating outside the optical trap,
\begin{equation} \label{eq.kc}
    k_c = k_c^{(0)} + i \frac{\Gamma m}{2\hbar k_c^{(0)}}.
\end{equation}
From experiment, we have estimated $k_c^{(0)} \approx 1.35$ $\mu$m$^{-1}$ by spatially filtering the polariton PL outside the pump spots. The imaginary term in Eq.~\eqref{eq.kc} describes the additional attenuation of polaritons due to their finite lifetime. We also account for coupling between the spins of the two condensates due to the TE-TM splitting which is captured with the parameter $\mathcal{J}$. The time delay parameter is approximated from the polariton phase velocity which gives,
\begin{equation}
    \tau = \frac{2dm}{\hbar k_c^{(0)}}.
\end{equation}
\begin{figure}
\centering
\includegraphics[width=\linewidth]{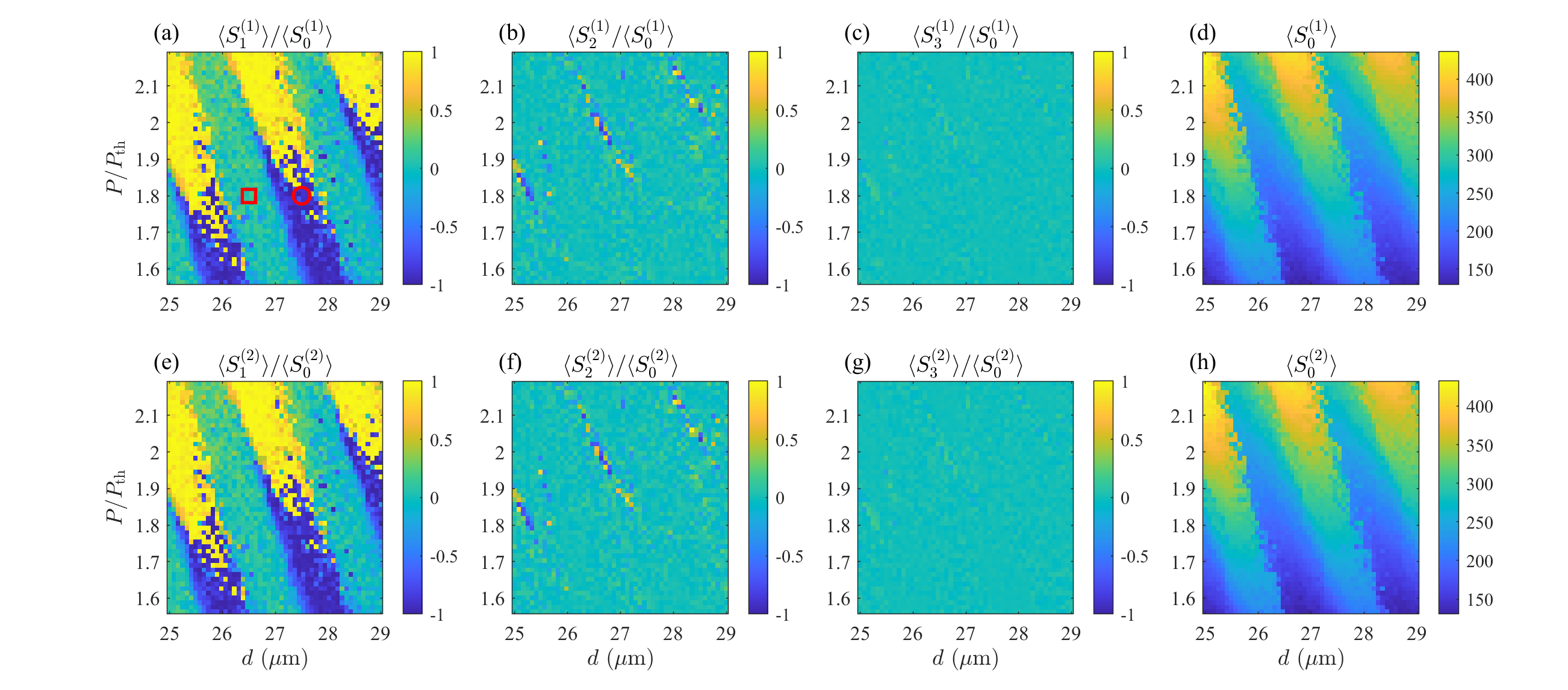}
\caption{(a-d) Time averaged Stokes parameters of $\psi_\pm^{(1)}$ and (e-h) $\psi_\pm^{(2)}$ from random initial conditions and varying pump power $P$ and distance $d$. We normalise the pump power in units of the threshold power for a single condensate $P_\text{th} = (\Gamma - 2 |\gamma|) \Gamma_R/R$. The parameters of the simulation are: $\epsilon=-0.01$ ps$^{-1}$; $\gamma = \epsilon/2$; $\Gamma = 0.25$ ps$^{-1}$; $\Gamma_R = \Gamma/4$; $\Gamma_s = \Gamma_R/2$; $\alpha = 0.15 \epsilon$; $R = 0.05\epsilon$; $\omega_0 = 5.5 \Gamma$; $g = \alpha$; $m = 0.3$ meV ps$^2$ $\mu$m$^{-2}$; $J_0 = 0.67 e^{1.8i}$ ps$^{-1}$; and $\mathcal{J} = 0.2 J$. Note that the large value of $J_0$ (as compared to $\epsilon$) is due to the smallness of the Hankel function in Eq.~\eqref{eq.hank}. At, e.g., $d=27$ $\mu$m we have $|J| = 2.64|\epsilon|$.}
\label{figS10}
\end{figure}
We show in Fig.~\ref{figS10} results of numerically integrating Eq.~\eqref{eq.orig2} from random initial conditions. Each pixel in the data is one realization of the condensate for the given power $P$ and distance $d$. The angled brackets of the Stokes parameters $\langle S_n^{(1,2)} \rangle$ represent time-average. We applied a constant step size Bogacki-Shampine algorithm~\cite{Flunkert2011} (a 3rd order Runge-Kutta). The timestep was chosen $\Delta t = 0.05$ ps and the integration was over $T = 5000$ ps for each condensate realization. The results reveal periodic polarization regimes similar to the phase-flip transitions recently reported in~\cite{Topfer_CommPhys2020}. At high powers we observe the condensates stabilizing into the ground state (horizontal) polarization [yellow colors in Figs.~\ref{figS10}(a,e)] whereas at low power we retrieve stable excited state (vertical) polarization condensation [blue colors in Figs.~\ref{figS10}(a,e)]. Between these bright yellow and blue regions we observe an intermediate region (seen as a mixture of blue and yellow datapoints) where the ground and excited state condensates are both stable and the random initial condition determines the winner. Such linear polarization bistability was already reported in~\cite{Sigurdsson_PRR2020s} for a single condensate. We also observe regions of complete depolarization (sea-green color) which correspond to condensate destabilization. Comparing Figs.~\ref{figS10}(a-d) with~\ref{figS10}(e-h) we observe, in the stable regime, that the condensates are strongly correlated in polarization (i.e., $\Psi^{(1)}$ and $\Psi^{(2)}$ always co-polarize). Our theoretical modeling gives good agreement with experimental observations presented in Fig.~5 in the main manuscript. In Fig.~\ref{figS11} we additionally show example dynamical trajectories from the unstable and stable regions marked by the red square and circle in Fig.~\ref{figS10}(a), respectively.
\begin{figure}[h]
\centering
\includegraphics[width=0.9\linewidth]{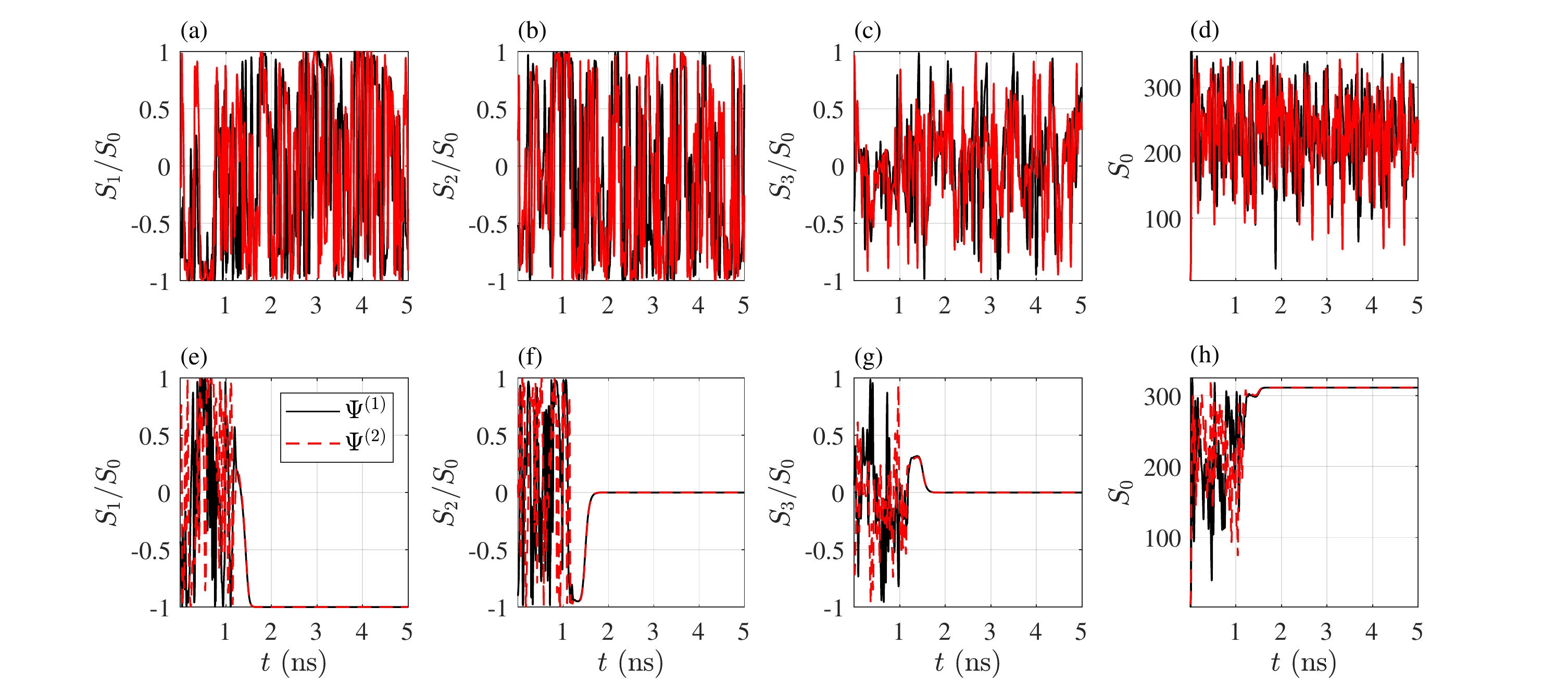}
\caption{Evolution of the Stokes parameters for a single realization of the condensate at the (a-d) the red square and (e-h) the red circle in Fig.~\ref{figS10}, corresponding to the unstable and stable regimes, respectively.}
\label{figS11}
\end{figure}

\section{Additional experimental data for coupled condensates}

\begin{figure}
\centering
\includegraphics[width=0.8\linewidth]{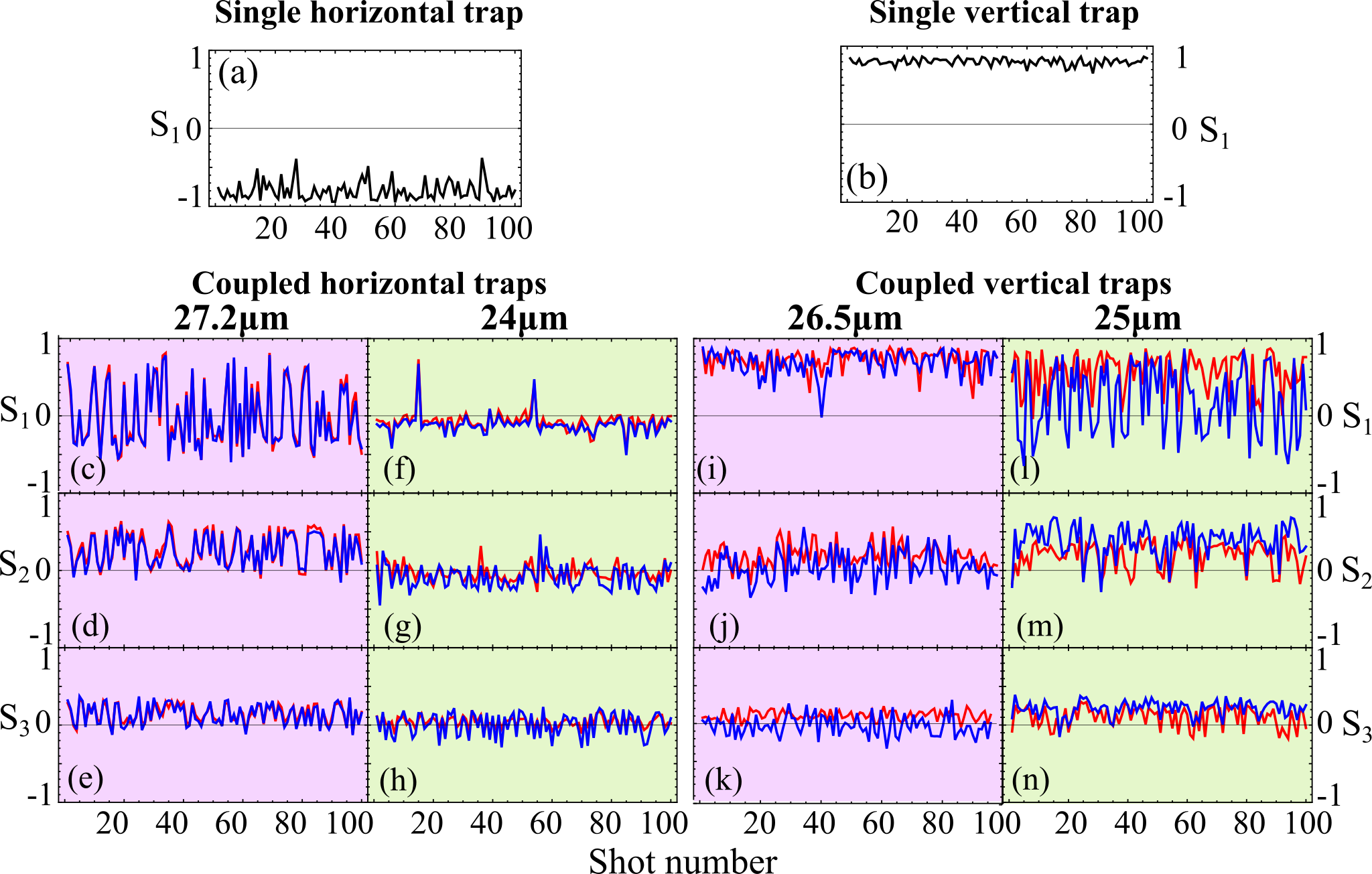}
\caption{100 realizations (shots) of the condensates Stokes components time-integrated in each 50 $\mu$s excitation shot. (a) and (b) shows $S_1$ of single horizontally and vertically elongated condensate (trap) respectively. $S_1$, $S_2$, and $S_3$ for two coupled condensates with their trap's major axes orientated longitudinally to the coupling direction and separated by 27.2 $\mu$m (c)-(e) and 24 $\mu$m (f)-(h), respectively. $S_1$, $S_2$, and $S_3$ for two coupled condensates with their trap's major axes orientated perpendicularly to the coupling direction, separated by 26.5 $\mu$m (i)-(k) and 25 $\mu$m (l)-(n), respectively. Blue and red colors correspond to right and left condensate respectively. Background green and purple colors depict different coupling regimes.}
\label{figcoupleexp}
\end{figure}

Here we present additional experimental data on the coupled elliptical condensates. In this experiment, we measure the Stokes components for 100 quasi-CW 50 $\mu$s shots. Each experimental point in Fig. \ref{figcoupleexp} represents the polarization component averaged within one excitation shot.
We note that the $S_1$, $S_2$ and $S_3$ in Figs.~ \ref{figcoupleexp}(c)-(n) are not measured simultaneously but consequently under the same pumping conditions and at the same position on the sample.

As expected, when isolated, the individual condensates stay dominantly polarized linearly parallel to the minor axis of the pump trap, as we describe in the main text [see Figs.~\ref{figcoupleexp}(a) and~\ref{figcoupleexp}(b) for a horizontal trap and vertical trap, respectively]. However, some fluctuations can be sometimes observed, for example, in the horizontally elongated trap in Fig.~\ref{figcoupleexp}(a). This happens due to some noise in our system as well due to mode competition. It is worth noting that such fluctuations decrease the values of the Stokes components presented in Figs.~2-4 in the main text since there we integrate/average over hundreds of shots.

In Figs.~\ref{figcoupleexp}(c)-(h) we present the Stokes components for coupled horizontally elongated ellipses separated by 27.2 $\mu$m  in Figs.~\ref{figcoupleexp}(c)-(e)  and 24 $\mu$m in Figs.~\ref{figcoupleexp}(f)-(h). Blue and red colors correspond to the "right" and "left" condensate, respectively. For 27.2 $\mu$m, the condensate flips randomly from horizontal to vertical polarization from shot-to-shot, whereas the $S_2$ has smaller values (less than 0.5) but also flips from shot to shot. Overall the $S_3$ component stays close to zero. For a different separation distance 24 $\mu$m shown in Figs.~\ref{figcoupleexp}(f)-(h) we observe that all Stokes components are close to zero in each shot. This means that the condensate pseudospin fluctuates rapidly in time within one excitation pulse with a zero mean polarization just like in simulation in Figs.~\ref{figS11}(a)-(d). Notice that the Stokes components still remain correlated indicating the condensate are coupled together.

We also plot all polarization components for two coupled vertically elongated condensates [Figs.~\ref{figcoupleexp}(i)-(n)]. The weaker coupling of such mutual trap configuration is evidenced through less correlations between the left and right condensates (i.e., the red and blue curves fluctuate more independently). For a distance of 26.5$\mu$m both condensates have strong horizontal polarization --- i.e. big $S_1$ component and small $S_2$ and $S_3$ components. This corresponds to Figs.~\ref{figS11}(e)-(h) in simulations. At a distance of 25 $\mu$m the condensates are in a semi-depolarized regime with oscillating $S_1$ and $S_2$ from shot to shot, and small $S_3$. 

\section{Power dependent pseudospin rotation}\label{sec.S10}
In this section we explain the results of Fig.~3(b) and~3(c) in the main manuscript where we can observe noticeable change in the linear polarization of the condensate as we increase the power. It manifests as counterclockwise rotation of the pseudospin in the equatorial plane of the Poincar\'{e} sphere.
     
\begin{figure}
\centering
\includegraphics[width=0.85\linewidth]{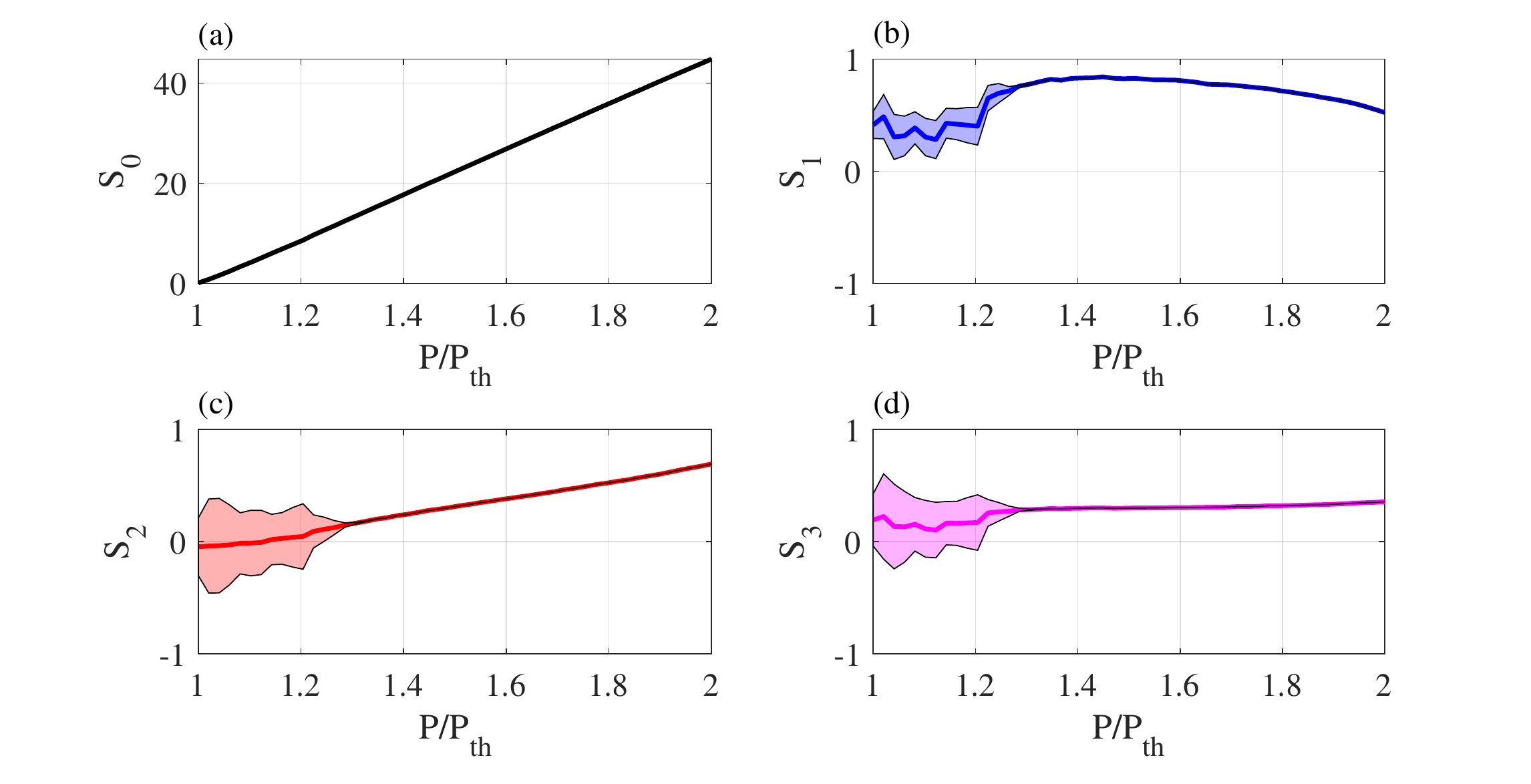}
\caption{Generalized Gross-Pitaevskii simulations of the condensate under spin-imbalanced pumping. (a)-(d) Time-integrated Stokes (pseudospin) components of the condensate for increasing pump power. Each datapoint is averaged over 100 random initial conditions. The shaded area is one standard deviation in the pseudospin dynamics (calculated over 5 ns) indicating nonstationary and stationary behaviour at low and high powers, respectively. Parameters are: $P_+/P_- = 1.0355$, $P = (P_++P_-)/2$, $P_\text{th} = \Gamma \Gamma_R/R$, $\Gamma = 1/5$ ps$^{-1}$, $\Gamma_R = 0.35 \Gamma$, $\Gamma_s = \Gamma_R$, $\epsilon = \Gamma/20$, $R = 0.015\Gamma$, $W = \Gamma_R$, and $g = R$.}
\label{fig.sim_ell}
\end{figure}

The explanation for the power dependent torque effect is due to slight polarization ellipticity in the excitation laser. This creates an imbalance between the spin-up and spin-down exciton populations in the system and a consequent out-of-plane effective magnetic field $\boldsymbol{\Omega}_\perp$ which rotates the pseudospin. This is confirmed through Gross-Pitaevskii simulations using Eq.~\eqref{eq.orig} where we introduce a slight pumping imbalance by redefining a spin-dependent pumping rate $P\to P_\pm$ where $P_+ \neq P_-$. We present our simulation in Fig.~\ref{fig.sim_ell} where we show the time-integrated Stokes (pseudospin) components of the condensate at increasing mean pump power [$P = (P_++P_-)/2$] where each datapoint is averaged over 100 random initial conditions. We have set $P_+/P_- = 1.0355$ and other parameters of the model (specified in the caption) are taken similar to the ones used in Fig.~\ref{figs3} and~\ref{figS10}. The shaded area is one standard deviation in the pseudospin dynamics (calculated over 5 ns) indicating nonstationary and stationary behaviour at low and high powers, respectively. We point out that Eq.~\eqref{eq.orig} must now include the additional pump induced blueshift $gP_\pm/W$ like in Eq.~\eqref{gpe} since it contributes to $\boldsymbol{\Omega}_\perp$. The results show precisely the counterclockwise rotation of the pseudospin in the $(S_1,S_2)$ plane like in Fig.~3(b) and~3(c) in the main manuscript. We have also confirmed that if $P_+ < P_-$ then the pseudospin rotates clockwise in the $(S_1,S_2)$ plane. 
    
Moreover, this change in the $S_1$ and $S_2$ can also be evidenced from the experimental data presented in Fig.~\ref{qwp}. There, a small change in the polarization ellipticity of our excitation beam dramatically affects the $S_1$ and $S_2$ distributions.

\end{document}